\documentclass[useAMS,usenatbib,english]{mn2e}
\usepackage[T1]{fontenc}
\usepackage[latin9]{inputenc}
\usepackage{verbatim}
\usepackage{amstext}
\usepackage{amssymb}
\usepackage{graphicx}
\usepackage{amsmath}
\usepackage{esint}
\usepackage{times}
\makeatletter

 \@ifundefined{textcolor}{}
 {%
   \definecolor{BLACK}{gray}{0}
   \definecolor{WHITE}{gray}{1}
   \definecolor{RED}{rgb}{1,0,0}
   \definecolor{GREEN}{rgb}{0,1,0}
   \definecolor{BLUE}{rgb}{0,0,1}
   \definecolor{CYAN}{cmyk}{1,0,0,0}
   \definecolor{MAGENTA}{cmyk}{0,1,0,0}
   \definecolor{YELLOW}{cmyk}{0,0,1,0}
 }

\usepackage{hyperref}
\hypersetup{
    colorlinks,%
    citecolor=blue,%
    filecolor=blue,%
    linkcolor=blue,%
    urlcolor=blue
}
\usepackage[all]{hypcap}

\providecommand{\eprint}[1]{\href{http://arxiv.org/abs/#1}{#1}}
\providecommand{\adsurl}[1]{\href{#1}{ADS}}
\providecommand{\ISBN}[1]{\href{http://cosmologist.info/ISBN/#1}{ISBN: #1}}

\newcommand{\citec}[1]{\bibpunct{(}{)}{,}{a}{}{,}\citet{#1}\bibpunct{(}{)}{;}{a}{}{,}}

\global\long\def\hDens{h^{3}\text{Mpc}^{-3}}
\global\long\def\hMpc{h^{-1}\text{Mpc}}
\global\long\def\hGpc{h^{-1}\text{Gpc}}
\global\long\def\by{\mathbf{y}}
\global\long\def\bx{\mathbf{x}}
\global\long\def\CR{\mathcal{R}}
\global\long\def\cD{\mathcal{D}}
\global\long\def\BE{\mathbb{E}}
\global\long\def\ol#1{\overline{#1}}
\global\long\def\av#1{\left\langle #1\right\rangle }
\global\long\def\rmd{{\rm d}}
\global\long\def\bk{\mathbf{k}}
\global\long\def\br{{\mathbf{r}}}
\global\long\def\CB{{\cal B}}
\global\long\def\cD{{\cal D}}
\global\long\def\CK{{\cal K}}
\global\long\def\d{\mathrm{d}}

\global\long\def\pasj{PASJ}
\global\long\def\prd{Phys. Rev. D}
\global\long\def\pre{Phys. Rev. E}

\global\long\def\apjs{ApJS}

\begin{document}
\title[Direct Minkowski Functionals of large redshift surveys]{Direct Minkowski Functional analysis of large redshift surveys:\protect\\ a new high--speed code tested on the luminous red galaxy Sloan Digital Sky Survey-DR7 catalogue}
\author[A. Wiegand, T. Buchert and M. Ostermann]{Alexander Wiegand$^{1}$, Thomas Buchert$^{2}$ and Matthias Ostermann$^{3,4}$\thanks{E--mail:alexander.wiegand@aei.mpg.de, buchert@obs.univ-lyon1.fr and mail@matthias-ostermann.de}\\
$^{1}$Max--Planck--Institut f\"ur Gravitationsphysik, Albert--Einstein--Institut,
 Am M\"uhlenberg 1, D--14476 Potsdam, Germany\\
$^2$Universit\'e de Lyon, Observatoire de Lyon, 
Centre de Recherche Astrophysique de Lyon, CNRS UMR 5574:\\ Universit\'e Lyon~1 and \'Ecole Normale Sup\'erieure de Lyon,
9 avenue Charles Andr\'e, F--69230 Saint--Genis--Laval, France\\
$^3$Arnold Sommerfeld Center, Ludwig--Maximilians--Universit\"at, Theresienstra\ss e 37, D--80333 M\"unchen, Germany\\
$^4$Oskar--Maria--Graf Gymnasium,
Keltenweg 5, D--85375 Neufahrn, Germany}

\date{Accepted 2014 June 4. Received 2014 May 12; in original form 2013 November 14}

\volume{443}
\pagerange{241--259} 
\pubyear{2014}

\label{firstpage}

\maketitle

\begin{abstract}
As deeper galaxy catalogues are soon to come, it becomes even more important
to measure large--scale fluctuations in the catalogues with robust statistics that cover all moments of the galaxy distribution.
In this paper we reinforce a direct analysis of galaxy data by employing the Germ--Grain method to calculate the
family of Minkowski Functionals. We introduce a new code, suitable for the analysis of large data sets without smoothing
and without the construction of excursion sets. We provide new tools to measure correlation properties, putting emphasis on
explicitly isolating non--Gaussian correlations with the help of integral--geometric relations. As a first application 
we present the analysis of large--scale fluctuations in the luminous red galaxy sample of Sloan Digital Sky Survey data release 7 data. We find
significant deviations from the $\Lambda$ cold dark matter mock catalogues on samples as large as $500\hMpc$ (more than $3\sigma$)
and slight deviations of around $2\sigma$ on $700\hMpc$, and we investigate possible sources of these deviations.
\end{abstract}


\section{Introduction}

Over the past decade, huge progress has been made in accessing the
galaxy distribution at larger and larger scales. At each step of this
process new and larger structures have been discovered, see e.g. \citec{2011ApJ...736...51E,2011EL.....9659001S,2012ApJ...759L...7P,2013MNRAS.429.2910C,2013ApJ...775...62K,2013arXiv1307.4405W}. However, to verify the reality of these structures, a robust statistical tool is mandatory \citep{2013MNRAS.434..398N,2013arXiv1310.2791N}.

The most common tool for the
characterization of large scale structure is based on two--point measures: the two--point correlation
function of the galaxy distribution and the complementary power spectrum.
They are particularly useful, as they can be related to the power
spectrum determined from the physics of the Early Universe. 
Claims that structures on scales of several hundreds
of megaparsecs are compatible with the $\Lambda$ cold dark matter (CDM) model are often
based on these lower order statistics.
However,
of course, they do not allow for a complete characterization of the
distribution.
This needs higher order
correlations that play an important role if the density field is not
Gaussian, especially when probing stages after
the formation of structure by gravitational collapse. Note, however, that also the full knowledge of all higher order correlations does not always characterize the distribution uniquely \citep{2012ApJ...750...28C}. As is well known, see e.g.~\citec{1993MNRAS.260..765C,1995A&A...294..345M}, 
a smoothed--out non--linear structure -- even if smoothed over very large scales -- is not described by structure described 
in linear gravitational instability, where in this latter, the distribution remains
Gaussian, if it was so in the initial data.

For this strongly clustered regime in the Late Universe, the Minkowski
Functionals that we are using in this paper provide a compact
and transparent framework to completely characterize the galaxy distribution.
As we shall see in Section~\ref{sec:Minkowski-functionals}, they
include all higher $N$--point correlations in a power series in the
sample density. We shall show in Section~\ref{sub:Importance-of-higher}
that it is indeed not enough to include only the lowest order contributions
of this series. This means that the values of the functionals that
we determine significantly depend on higher order clustering. As is also  well known,
only higher order correlations are sensitive to the morphology
of large--scale structure.

Due to this interesting property of including higher correlations
in a simple way, the Minkowski Functionals have been determined for
many galaxy and cluster surveys. The specific Germ--Grain model that
has been introduced into cosmology together with the family of Minkowski
Functionals in \cite{1994A&A...288..697M,1995lssu.conf..156B} (see
\cite{1996dmu..conf..281S} for a brief tutorial), and which will
be briefly reviewed in Section~\ref{sub:Boolean-Grain-model}, has
been used for example for the IRAS 1.2Jy and PSCz surveys \cite{1996app..conf...83K,1998A&A...333....1K,2001A&A...373....1K},
and the Abell/ACO cluster catalogue \citep{1997MNRAS.284...73K}. For
an early sample of the SDSS catalogue (data release DR 3), the Minkowski
Functionals have been determined for smoothed isodensity contours
of the galaxy distribution in \cite{2003PASJ...55..911H}, which is
also the method employed in most work on Minkowski Functional analysis
of galaxy and cluster catalogues as well as in simulations 
\cite{1996app..conf..251P,1997ApJ...482L...1S,1998ApJ...495L...5S,1998ApJ...508..551S,1999ApJ...526..568S,2004astro.ph..8428N,2013arXiv1310.6810B,2013ApJS..209...19C}.
A further, recently proposed smoothing technique directly uses the Delauney tessellation of
the point distribution \citep{2010arXiv1006.4178A}.

To robustly measure non--Gaussianity with the help of Minkowski Functionals is mostly discussed for isotemperature contour maps of the cosmic microwave background
\cite{1998MNRAS.297..355S,2000ApJ...544L..83S,2012MNRAS.425.2187H,2013MNRAS.429.2104D,2013MNRAS.428..551M,2013MNRAS.434.2830M,2013arXiv1303.5083P}, as well as for studies of the weak lensing convergence field \citep{2012PhRvD..85j3513K,2013arXiv1309.4460P}.

One focus of this paper is the determination of non-Gaussian features from Minkowski Functionals in three--dimensional galaxy data, which has been addressed in \citet{2012MNRAS.423.3209P} and \citet{2013MNRAS.435..531C}. The other focus lies on reinforcing
the Germ--Grain method in the three dimensional case to calculate the Minkowski Functionals. We shall apply this
method to the SDSS DR 7. This release was chosen due to its complete angular coverage of the SDSS survey region and the existence of a detailed standard analysis of the two--point correlation properties in \citet{2010ApJ...710.1444K}.
In order to be able to probe larger scales than before in \citet{2003PASJ...55..911H}, we specifically use the luminous red galaxy (LRG) sample
of the DR 7 in the compilation of \citet{2010ApJ...710.1444K}.
Newer and upcoming data will be analysed in forthcoming work. Especially the full SDSS DR 12 catalogue, but also catalogues of after--Sloan programmes, are targets for our optimized code.

The paper is organized as follows: Section~\ref{sec:Minkowski-functionals}
recalls basic properties of the Minkowski Functionals and briefly
reviews the Germ--Grain model for the direct analysis of the galaxy
point process. We discuss the usefulness of this model by collecting
the analytical results that are known for the Minkowski Functionals
of this model, examine the Gauss--Poisson process, and introduce our method to extract information
on higher order correlations from the Minkowski Functionals of the
model. Section~\ref{sec:code} describes the new code
that we use in order to efficiently calculate the Minkowski Functionals
in the Germ--Grain model for a large data set like the SDSS LRG catalogue.
Section~\ref{sec:Minkowski-fun-LRG} presents and discusses the results for two
different subsamples at different luminosity thresholds of this catalogue as a first
application of our methods.
In Section~\ref{sec:Non-Gaussian-correlations},
we derive the values of some integrals over the two-- and three--point
correlation function and study the deviations of the point distribution
from a Gauss--Poisson process and also from a lognormal distribution. We here explicitly demonstrate that the low--order correlations
are actually not enough to describe the structure in the data set.
We conclude in Section~\ref{sec:Conclusion}.

\section{Minkowski Functionals of the Germ--Grain model}
\label{sec:Minkowski-functionals}

Let us begin by a description of Minkowski Functionals in the Germ--Grain
model that we shall use here. For a more complete description
see \cite{mecke2000statistical}, \cite{phdSchmalzing}, \citet{beisbart2001measuring} or \cite{UBHD1591152}.

\begin{figure}
\includegraphics[width=1\linewidth]{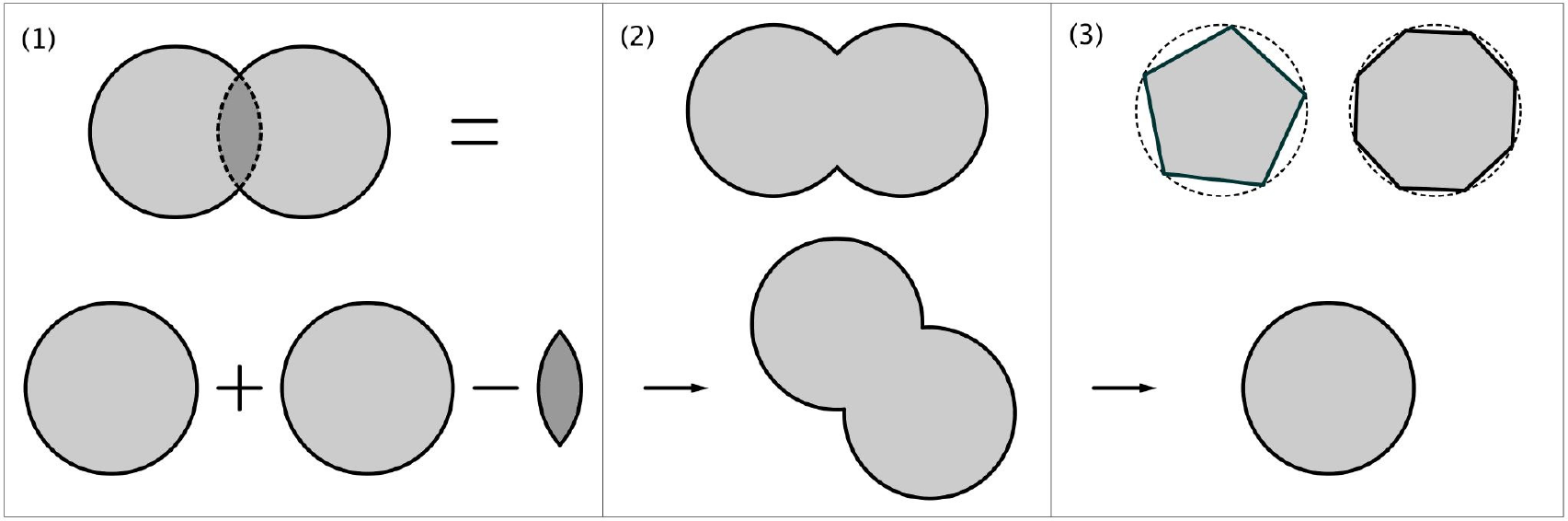}
\caption[Properties of the scalar functionals]{Properties of the scalar functionals: (1) Additivity, (2) Motion
invariance, (3) Conditional continuity.}
\label{img:properties}
\end{figure}

\subsection{Minkowski Functionals}

Minkowski Functionals are morphological descriptors of extended bodies that rely on
well--developed results in integral geometry. In 3D Euclidean space there
are four of them that we shall label $V_{0}-V_{3}$. In the normalization
we use, they are related to geometrical properties of the body as
follows: 
\begin{equation}
V_{0}=V\;\;;\;\; V_{1}=\frac{S}{6}\;\;;\;\; V_{2}=\frac{H}{3\pi}\;\;;\;\; V_{3}=\chi\;.\label{eq:MinkRelGeo}
\end{equation}
Here, $V$ is the volume of the body, $S$ is its surface area, $H$
the integral mean curvature of the surface and $\chi$ the Euler characteriztic (the integral Gaussian curvature of the surface).

The surprising fact, shown in \cite{UBHD152758}, is that every other
scalar functional that can be defined to describe a given body and
that fulfils the properties of motion invariance, additivity and
conditional continuity (sketched in Fig.~\ref{img:properties}), can
be expressed as a linear combination of the four functionals of Equation
(\ref{eq:MinkRelGeo}).

Instead of working with the functionals $V_{\mu}$, we will more often need the corresponding densities $v_{\mu}$. They are simply defined by
\begin{equation}
v_{\mu}=V_{\mu}/\left|\cD\right|\;,
\end{equation}
where $\left|\cD\right|$ is the volume within the survey boundary.

\subsection{Germ--Grain model\label{sub:Boolean-Grain-model}}

The Minkowski Functionals as described in the previous section are only
defined for extended bodies. To use them for characterizing the galaxy
distribution, one has to define a procedure that transforms the point
distribution into a collection of extended objects. The two major
methods that are used so far to achieve this are the construction of {\em excursion
sets} (e.g. isodensity contours) and the {\em Germ--Grain model}.

In order to determine isodensity contours, the point particle distribution
is smoothed into a continuous density field. The surfaces of
a given density threshold then provide the boundaries of the body that
we are going to analyse. The values of the four Minkowski Functionals
(volume, surface area, mean curvature, Euler characteriztic) can then be
determined as a function of the (over)density that is used to determine
the isodensity contours \citep{1997ApJ...482L...1S}. This method is commonly employed in the
community (see the reference list in the introduction), and it has
also been used for the SDSS data in \cite{2003PASJ...55..911H}.

\begin{figure}
\centering 
\includegraphics[width=0.9\linewidth]{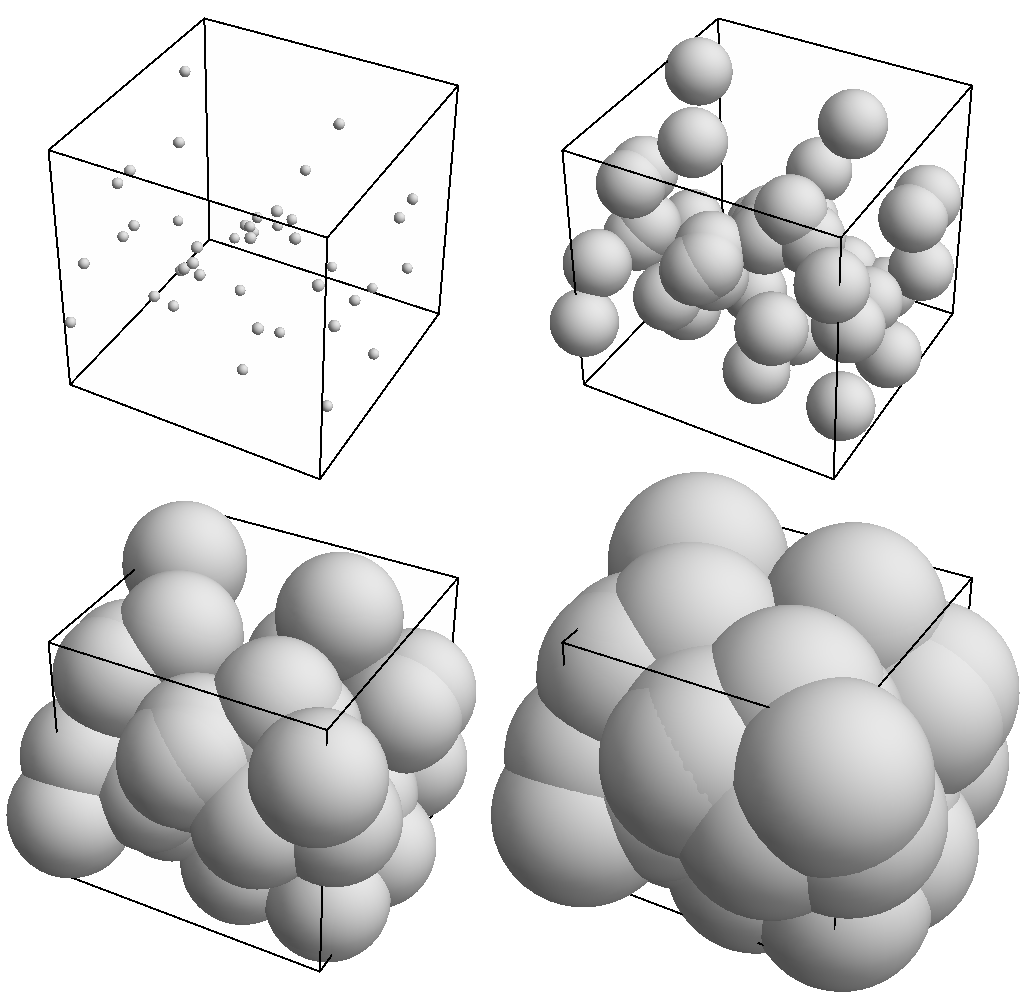}
\caption[Germ Grain model]{By increasing the radius of Balls around the points $\{\mathbf{x}_{1},\dots,\mathbf{x}_{N}\}$
up to the maximum radius $R$, more and more Balls intersect and a
complex structure develops.}
\label{img:boolean} 
\end{figure}

In the Germ--Grain model, the point distribution of galaxies is converted
into a set of extended bodies by decorating each galaxy with a Grain
(here a Ball of radius $R$, but any shape of Grain could in principle be used
to take into account internal morphologies). Instead of the (over)density, the
(equal) radius of these Balls can be used as a diagnostic parameter,
i.e. to present the values of the four functionals as a function of
scale \citep[the radius $R$ of the Balls;][]{1994A&A...288..697M}.  This results in quite complex structures as shown in Fig.~\ref{img:boolean}. 
Also this method has been used quite extensively in the past to characterize galaxy and cluster distributions,
e.g. in \citec{1996app..conf...83K,1997MNRAS.284...73K,1998A&A...333....1K,2001A&A...373....1K,1996app..conf..251P} and for other statistics in \citet{1999ApJ...513..543K},
but it has not become a standard analysis tool in cosmology. 
With this paper we emphasize the advantages of this direct method to analyse galaxy data.

Comparing the two methods, the Germ--Grain model has several important
advantages over the construction of excursion sets: 
\begin{enumerate}
\item it can be implemented in an easy and robust manner. This simplicity
also implies, that 
\item we have an analytical understanding of the relation of the average Minkowski
Functional (densities) to the connected correlation functions of the
underlying point distribution, through which also the global contribution of Poisson noise is explicitly known;
\item the global functionals are represented by their local contributions
(so--called {\em Partial Minkowski Functionals}). This local information can be
used to extract subsamples with given environmental properties. The
partial functionals form the basis of image analysis techniques, since
they allow us to extract filamentary or cluster galaxies from a distribution,
even if these morphological properties are strongly diffused by Poisson noise
\citep{1996app..conf..251P}\footnote{See also \citet{2008JSMTE..12..015M} for a more recent application of Minkowski Functionals in image analysis.}. 
\end{enumerate}
These useful relations, e.g. between the connected correlation functions
and the average Minkowski Functional densities, as well as the image analysis
properties of partial functionals have been demonstrated in \cite{phdSchmalzing}.

\subsection{Statistical interpretation\label{sub:stat}}

Thus far, the Germ--Grain Minkowski Functionals were mainly used for a comparison of individual samples. These comparisons made use of the property of the Minkowski Functionals to provide a morphological characterization of the galaxy distribution in the analysed sample: if the Germ--Grain Minkowski Functionals differ for two samples, these are morphologically distinct \citep{1998A&A...333....1K}.

In addition to this comparison of individual point sets, we here also want to extract some information on the statistical properties of the point distribution that underlie these individual galaxy data sets.
We, therefore, interpret an observed or simulated galaxy sample, as usual, as a particular realization of a point process with a priori unknown statistical properties. This gives rise to an ensemble of Minkowski Functionals $\left\{V_{\mu}\right\}$ or their corresponding densities $\left\{v_{\mu}\right\}$, respectively. As made explicit in Appendix~\ref{sec:derivation}, it can be shown that the ensemble average of the densities $v_{\mu}$ is related to the connected correlation functions $\xi_{n}$ of the point process as follows:
\begin{eqnarray}
\av{v_{0}} & = & 1-e^{-\varrho_{0}\overline{V}_{0}}\;\nonumber \\
\av{v_{1}} & = & \varrho_{0}\overline{V}_{1}e^{-\varrho_{0}\overline{V}_{0}}\;,\nonumber \\
\av{v_{2}} & = & \left(\varrho_{0}\overline{V}_{2}-\frac{3\pi}{8}\varrho_{0}^{2}\overline{V}_{1}^{2}\right)e^{-\varrho_{0}\overline{V}_{0}}\;,\nonumber \\
\av{v_{3}} & = & \left(\varrho_{0}\overline{V}_{3}-\frac{9}{2}\varrho_{0}^{2}\overline{V}_{1}\overline{V}_{2}+\frac{9\pi}{16}\varrho_{0}^{3}\overline{V}_{1}^{3}\right)e^{-\varrho_{0}\overline{V}_{0}}\;.\label{eq:MinkDensDef}
\end{eqnarray}
For a Poisson distribution, the quantities $\overline{V}_{\mu}$ are simply
the Minkowski Functionals $V_{\mu}\left(B\right)$ of the Balls of common radius $R$ that
we use to obtain extended bodies:
\begin{eqnarray}
&&V_{0}\left(B\right)=\frac{4\pi}{3}R^{3}\;\;;\;\; V_{1}\left(B\right)=\frac{2}{3}\pi R^{2}\;\;;\nonumber\\
&&V_{2}\left(B\right)=\frac{4}{3}R\;\;;\;\; V_{3}\left(B\right)=1\;.\label{eq:MinkBall}
\end{eqnarray}
For a point distribution with structure, the $\overline{V}_{\mu}$ 
pick up contributions that depend on the dimensionless connected correlation function of order
$n+1$, $\xi_{n+1}$, as 
\begin{eqnarray}
\overline{V}_{\mu} & = & V_{\mu}\left(B\right)+\sum_{n=1}^{\infty}\frac{\left(-\varrho_{0}\right)^{n}}{\left(n+1\right)!}\int_{\cD}\d^{3}x_{1}\dots\d^{3}x_{n}\label{eq:MinkCorrCon}\\
 &  & \times\xi_{n+1}\left(0,\bx_{1},\dots\bx_{n}\right)V_{\mu}\left(B\cap B_{\bx_{1}}\cap\dots\cap B_{\bx_{n}}\right)\;.\nonumber 
\end{eqnarray}
The integrals run over the positions of the centres $\bx_{i}$ of
the Balls $B_{\bx_{i}}$. As it is the intersection of all Balls $B\cap B_{\bx_{1}}\cap\dots\cap B_{\bx_{n}}$
that enters, the integrals vanish for configurations where the $\bx_{i}$
are separated by more than $2R$. Therefore, determining the Minkowski
Functionals as a function of the Ball radius $R$ probes the correlation
of the point distribution up to a scale of $2R$. We shall exploit
this property in Section~\ref{sec:Non-Gaussian-correlations}.

The introduction of the dimensionless version of the connected correlation functions $\xi_{n}$ requires the assumption that the global point process possesses a well--defined non--zero and stable (scale--independent) average density $\varrho_{0}=\av{\varrho(r)}$, a requirement that is expected to hold for an existing homogeneity scale.
Note, however, that this assumption is not required for the Minkowski Functional analysis itself. Also for expressing the $\av{v_{\mu}}$ in terms of the statistical quantities describing the point process, one could work with the dimensionfull connected correlation functions, without assuming that $\varrho_{0}\neq0$. We shall only need this condition for the extraction procedure described in Section~\ref{sub:Extracting-higher-order}. The reasoning in that case is then, that we interpret the analysed sample as being a representative realization of the full point process. We assume that it has a positive average density and estimate this background density $\varrho_{0}$ from the sample.
Of course, this regional estimate can be biased relative to the true global value (assumed to exist): for an analysis of the correlation properties well inside the survey region, we consider this assumption as sufficiently accurate, since the estimation of the correlation properties would be most strongly influenced on the scale of the sample (on this scale the integral constraint requires vanishing of the fluctuations). This subtlety is worth to be kept in mind in future analyses.

\subsection{Gauss--Poisson process\label{sub:Gauss-Poisson-process}}

To get a better intuition about the influence of correlation functions
on the modified Minkowski Functionals $\overline{V}_{\mu}$, we shall
first consider the case of a Gaussian distribution. For low enough
average density $\varrho_{0}$ and certain correlations $\xi_{2}$,
a point distribution can be described by a Gauss--Poisson process.
\citet{2001PhRvE..64e6109K} shows that for this to be possible, the correlation function 
has to be non--negative, $\xi_{2}\ge0$,
and the average density $\varrho_{0}$ has to satisfy
\begin{equation}
\varrho_{0}\int_{\cD}d\by\xi_{2}\left(\left|\by\right|\right)\le1\;,
\label{eq:xicond}
\end{equation}
on the domain $\cD$ of the point sample.

The resulting Gauss--Poisson process has the property that all
higher connected correlation functions $\xi_{n}$ for $n>2$ are zero.
This drastically simplifies the expressions for the $\overline{V}_{\mu}$.
Equation~(\ref{eq:MinkCorrCon}) becomes 
\begin{equation}
\overline{V}_{\mu}=V_{\mu}\left(B\right)-\frac{\varrho_{0}}{2}\int_{\cD}\d^{3}x_{1}\xi_{2}\left(\left|\bx_{1}\right|\right)V_{\mu}\left(B\cap B_{\bx_{1}}\right)\;.\label{eq:MinkGaussExp}
\end{equation}
 We already gave the explicit expressions for the Minkowski Functionals
of the Balls $V_{\mu}\left(B\right)$ in Equation~(\ref{eq:MinkBall}).
For a known correlation function, it is in addition possible to calculate
the contribution of the second term. To this end, we have to determine
the Minkowski Functionals for the intersection of two Balls $V_{\mu}\left(B\cap B_{\bx_{1}}\right)$.
This intersection has the form of a convex lens and it is easy to
figure out its volume and surface as a function of the distance $r$,
separating the centres of the Balls. For the mean curvature, the result
can be found in \citet{UBHD1591152}. The form of the $V_{\mu}\left(B\cap B_{\bx_{1}}\right)$
in a spherical coordinate system centred on $B$ is then the following:
\begin{eqnarray}
V_{0}\left(r\right) & = & \frac{1}{12}\pi\left(2R-r\right)^{2}\left(r+4R\right)\;\;;\label{eq:V0r}\\
V_{1}\left(r\right) & = & \frac{1}{3}\pi R\left(2R-r\right)\;\;;\\
V_{2}\left(r\right) & = & \frac{2}{3}\left(2R-r\right)+\frac{2}{3}R\sqrt{1-\left(\frac{r}{2R}\right)^{2}}\arcsin\left(\frac{r}{2R}\right);\label{eq:Mink2TwoBall}\\
V_{3}\left(r\right) & = & 1\;\;,\label{eq:V3r}
\end{eqnarray}
where $R$ is again the radius of the Balls. 
As they do not intersect if the separation of the two centres is
larger than $2R$, $V_{\mu}\left(B\cap B_{\bx_{1}}\right)=0$ for
$r>2R$. Therefore, the integral~(\ref{eq:MinkGaussExp}) reduces to
\begin{equation}
\overline{V}_{\mu}=V_{\mu}\left(B\right)-2\pi\varrho_{0}\intop_{0}^{2R}V_{\mu}\left(r\right)\xi_{2}\left(r\right)r^{2}\d r\;.\label{eq:MinkGaussExpRform}
\end{equation}

\begin{figure}
\includegraphics[width=0.47\textwidth]{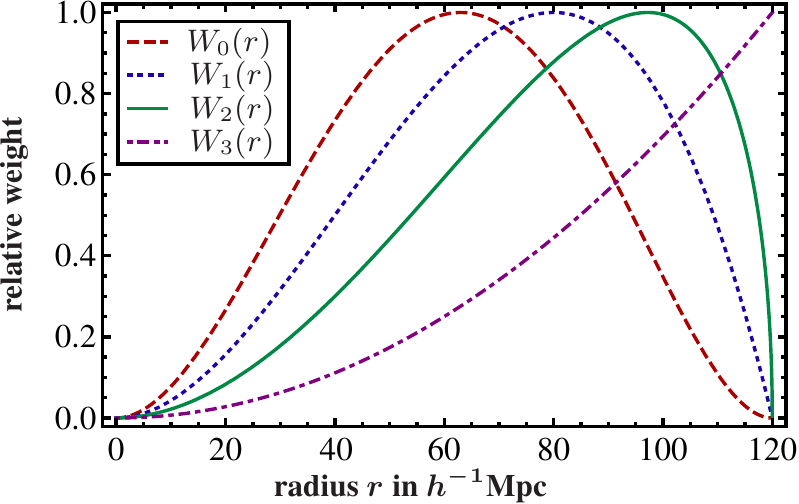}
\caption{Form of the integration windows $W_{\mu}\left(r\right)=V_{\mu}\left(r\right)r^{2}$ in Equations
(\ref{eq:V0r})-(\ref{eq:V3r}) used in the integral~(\ref{eq:MinkGaussExpRform}), for a Ball radius of $R=60\hMpc$.
The functionals of higher index $\mu$ probe the correlation function
at larger distances.}
\label{fig:IntegrationWindow}
\end{figure}
To get a feeling what aspects of the correlation function these integrals
probe, we show the form of the window functions $V_{\mu}\left(r\right)r^{2}$
in Fig.~\ref{fig:IntegrationWindow}.

To calculate these integrals
for the standard $\Lambda$CDM structure, it is useful to express
them directly in terms of the power spectrum. Inserting the Fourier
transform of $\xi_{2}\left(r\right)$ gives 
\begin{equation}
\overline{V}_{\mu}=V_{\mu}\left(B\right)-\frac{\varrho_{0}}{\pi}\intop_{0}^{\infty}P\left(k\right)W_{\mu}\left(k,R\right)k^{2}\d k \;\;,\label{eq:TheoryVmu}
\end{equation}
 with the functions 
\begin{eqnarray}
W_{0}\left(k,R\right) & = & \frac{4\pi(\sin(kR)-kR\cos(kR))^{2}}{k^{6}}\;;\\
W_{1}\left(k,R\right) & = & -\frac{2\pi R(kR\sin(2kR)+\cos(2kR)-1)}{3k^{4}}\;;\\
W_{2}\left(k,R\right) & = & -\frac{4(kR\sin(2kR)+\cos(2kR)-1)}{3k^{4}}\nonumber \\
 &  & +\frac{2}{3}\frac{R}{k}\intop_{0}^{\pi/2}\sin(\phi)\sin(2kR\sin(\phi))\phi d\phi\;; \\
W_{3}\left(k,R\right) & = & \frac{\sin(2kR)-2kR\cos(2kR)}{k^{3}}\;.
\end{eqnarray}
As $W_{0}\left(k,R\right)$ is simply the square of the Fourier transform
of a top hat window function, $\overline{V}_{0}$ can be related to
another well--known statistical property of a point distribution,
namely the matter variance in a sphere of radius $R$: 
\begin{equation}
\sigma^{2}\left(R\right)=\frac{1}{\left(2\pi\right)^{3}}\int \d^{3}kP\left(\bk\right)\left|\widetilde{W}_{\CB\left(R\right)}\left(k\right)\right|^{2}\;.\label{eq:sigmaPower}
\end{equation}
This means that $\overline{V}_{0}$ and $\overline{V}_{3}$ are
directly related to the two--point quantities by 
\begin{equation}
\overline{V}_{0}=\frac{4\pi}{3}R^{3}\left(1-\frac{\frac{4\pi}{3}R^{3}\varrho_{0}}{2}\sigma^{2}\left(R\right)\right)\label{eq:V0-sigma}\;,
\end{equation}
and 
\begin{equation}
\overline{V}_{3}=1-2\pi\varrho_{0}\intop_{0}^{2R}\xi_{2}\left(r\right)r^{2}\d r\;.\label{eq:V3-xi}
\end{equation}
The other two modified functionals $\overline{V}_{1}$, $\overline{V}_{2}$ are not as directly related to other well--known quantities, but
probe additional aspects of the form of the correlation
function defined by their weights in Fig.~\ref{fig:IntegrationWindow}.

\subsection{Extracting higher order correlations\label{sub:Extracting-higher-order}}

As we have seen in the previous section, we can directly derive certain
integrals over the two--point correlation function from the Minkowski
Functionals of a Gauss--Poisson point distribution. For a more general
point distribution this is no longer straightforward, but we can use
the fact that we know the exact form of the dependence of $\av{v_{\mu}}$
on the average density of the point process $\varrho_{0}$.
For a single sample, of course, this density is fixed. By random subsampling of the original sample, however, we can create (noisier) samples of a lower average density.
In this way, we can determine the $\av{v_{\mu}}$
not only as a function of the radius of the Balls, but also as a function
of the density of the point distribution. This allows us to extract the integrals
over the correlation functions in~(\ref{eq:MinkCorrCon}) as follows.

Let us assume that we are able to measure the $\av{v_{\mu}}$ accurately
for a given density. By repeating this measurement for several
densities $\varrho_{0}$, we get an approximation to the functional
dependence of $v_{\mu}$ on $\varrho_{0}$. Then, by inverting the
system~(\ref{eq:MinkDensDef}), we can derive from the measured values
of $\av{v_{\mu}\left(\varrho_{0}\right)}$ the corresponding functional
dependence of $\overline{V}_{\mu}$ on $\varrho_{0}$. Calling this
empirical function $\tilde{\overline{V}}_{\mu}\left(\varrho_{0}\right)$,
we know that it can be written as a series expansion in $\varrho_{0}$
of precisely the form~(\ref{eq:MinkCorrCon}). This means that, if
we can compute the coefficients of this series, we shall obtain the
corresponding weighted integrals over the correlation functions. Taylor expanding
$\tilde{\overline{V}}_{\mu}\left(\varrho_{0}\right)$ around $\varrho_{0}=0$,
we obtain
\[
\tilde{\overline{V}}_{\mu}\left(\varrho_{0}\right)=\sum_{n=0}^{\infty}\frac{\tilde{\overline{V}}_{\mu}^{\left(n\right)}\left(0\right)}{n!}\varrho_{0}^{n}\;\;,
\]
 where the exponent $\left(n\right)$ stands for the $n$--th derivative
of $\tilde{\overline{V}}_{\mu}\left(\varrho_{0}\right)$ with respect
to $\varrho_{0}$. These derivatives can now be directly related to
the coefficients of the expansion~(\ref{eq:MinkCorrCon}). Writing
this expansion in short as 
\[
\overline{V}_{\mu}=\sum_{n=0}^{\infty}\frac{b_{n+1}^{\mu}}{\left(n+1\right)!}\left(-\varrho_{0}\right)^{n}\;\;,
\]
with $b_{1}=V_{\mu}\left(B\right)$, we deduce that
\begin{eqnarray}
b_{n+1}^{\mu} & = & \int_{\cD}\xi_{n+1}\left(0,\bx_{1},\dots\bx_{n}\right)V_{\mu}\left(B\cap B_{\bx_{1}}\cap\dots\cap B_{\bx_{n}}\right)\nonumber\\
 &  & \times\d^{3}x_{1}\dots\d^{3}x_{n}
  =  \left(n+1\right)\left(-1\right)^{n}\tilde{\overline{V}}_{\mu}^{\left(n\right)}\left(0\right)\;\;.\label{eq:MinkIntegr}
\end{eqnarray}
This implies that we are able to quantitatively determine, how much
the point distribution deviates from a pure Gauss--Poisson distribution.
The `Gaussian part' is related to the first derivative of $\tilde{\overline{V}}_{\mu}\left(\varrho_{0}\right)$
at zero density, and it especially allows us to compare the result to other
independent measurements of $\sigma^{2}\left(R\right)$ and $\xi_{2}\left(R\right)$
via the relations~(\ref{eq:V0-sigma}) and~(\ref{eq:V3-xi}).

To carry out this procedure in practice, we estimate $\av{v_{\mu}}$ by the Minkowski Functional density $v_{\mu}$ of the given realization of the point process.
For densities lower than the original density, we average $v_{\mu}$ of several random subsamples.
To determine how accurate this estimate is, we use the average and fluctuations of the individual $v_{\mu}$ in an ensemble of mock samples produced from simulations.
This allows us to test, whether the observed sample is consistent with the simulated cosmology.

For the three--point functions, the quantity with the simplest weight
functions in the integral reads: 
\begin{eqnarray}
b_{3}^{0} & = & \int_{\cD}\rmd^{3}x_{1}\rmd^{3}x_{2}\rmd^{3}x_{3}\zeta\left(\left|\bx_{1}-\bx_{3}\right|,\left|\bx_{2}-\bx_{3}\right|,\left|\bx_{1}-\bx_{2}\right|\right)\times\nonumber\\
 &  & \times\theta\left(R-\left|\bx_{1}\right|\right)\theta\left(R-\left|\bx_{2}\right|\right)\theta\left(R-\left|\bx_{3}\right|\right)\;.\label{eq:ThreePtInteg}
\end{eqnarray}
 The other integrals of the three--point function are more complicated
and we would not write them out explicitly. We calculate them numerically, using our code to determine $V_{\mu}\left(B\cap B_{\bx_{1}}\cap B_{\bx_{2}}\right)$ for the cases where $\mu\neq0$.

To characterize the deviation of the galaxy distribution in the SDSS
from a Poisson and a Gauss--Poisson distribution, we shall determine
the coefficients $b_{n+1}^{\mu}$ in Section~\ref{sec:Non-Gaussian-correlations}
for $n\leq2$.

\section{The new code package: \textsc{\protect\footnotesize MINKOWSKI--4}}
\label{sec:code}
 
As described in the previous section, we can learn a lot about the
structure in the galaxy distribution in the Universe and especially about the magnitude of higher order
correlations, if we are able to calculate the Minkowski Functional
densities $v_{\mu}$ accurately.

With this paper we provide the \textsc{\footnotesize MINKOWSKI--4} package, built on the new code \textsc{\footnotesize CHIPMINK} (Code for High--speed Investigation of Partial Minkowski Functionals), which is a completely revised version of a code based on previous
work by Jens Schmalzing and Andreas Rabus in 1998, see \cite{rabus-dipl}.
The package compiles modules to compute the Minkowski Functionals of a given point sample
for the Germ--Grain model [which generalises the Boolean Grain model -- where the
Germs are those of a Poisson process \citep{stoyan1987stochastic} -- to arbitrary
point distributions]. It extracts correlation properties of the point set in the form of the Minkowski functional densities~(\ref{eq:MinkDensDef}) and the modified Minkowski Functionals $\overline{V}_{\mu}$. Optionally, it also delivers the full set of Partial Minkowski Functionals of the  environmental morphology of every point in the sample.

\subsection{Computation of the Germ--Grain model}

The computational methods for the Germ--Grain model of the Minkowski Functionals (henceforth abbreviated as MFs) heavily rely
on the work of Mecke, Buchert and Wagner in \cite{1994A&A...288..697M}
and are therefore also strongly related to the works of Kerscher and
collaborators \citep{1997MNRAS.284...73K,1998A&A...333....1K,2001A&A...373....1K,2001A&A...377....1K}.

As outlined in Section~\ref{sub:Boolean-Grain-model}, a sphere of
radius $r$ (the so--called Grain or Ball) is placed around each point
of the sample, the Germ. The union of the Balls then forms the structure
$\CB_{r}$, 
\begin{equation}
\CB_{r}=\bigcup_{i=1}^{N}\CB(\mathbf{x}_{i};r)\;.
\end{equation}
When we increase the Balls' radius $r$ up to a maximum radius $R$,
a more and more complex structure develops, see Fig.~\ref{img:boolean}.
Thus, the radius $0\leq r\leq R$ serves as a diagnostic parameter.

In the Germ--Grain model, the global MFs -- apart from the volume --
are localized on the surface of the structure and can be determined
by means of the so--called partition formula (see e.g.~\cite{2000MNRAS.312..638S}),
\begin{equation}
V_{\mu}(\CB_{r})=\sum_{i=1}^{N}V_{\mu}^{(i)}+\frac{1}{2}\sum_{i,j=1}^{N}V_{\mu}^{(ij)}+\frac{1}{6}\sum_{i,j,k=1}^{N}V_{\mu}^{(ijk)}\;,\label{partition-formula}
\end{equation}
 where $V_{\mu}^{(i)}$ is the contribution of the Ball around $\mathbf{x}_{i}$
(at given radius $r$), and where $V_{\mu}^{(ij)}$ and $V_{\mu}^{(ijk)}$
are those of its intersection with one, respectively, two neighbours.

\subsection{Partial MFs}

The global MFs can be calculated by adding up Partial MFs assigned
to each Grain, see for example \cite{1994A&A...288..697M} and in
an application \cite{2000MNRAS.312..638S}, 
\begin{equation}
V_{\mu}(\CB_{r})=\sum_{i=1}^{N}V_{\mu}(\mathbf{x}_{i};r)\;, \label{eq:sumpart}
\end{equation}
where $V_{\mu}(\mathbf{x}_{i};r)$ are the Partial MFs of the Ball
around $\mathbf{x}_{i}$ with radius $r$. These can be determined
by the local intersections of the Balls. Since only neighbours within
$2r$ around a point contribute to its Partial MFs, we determine a
\emph{neighbourlist} for each point of the sample before the actual
calculation, which consists of the points within a distance of two
times the maximum radius (as well as the point itself).

\begin{figure}
\includegraphics[width=0.47\linewidth]{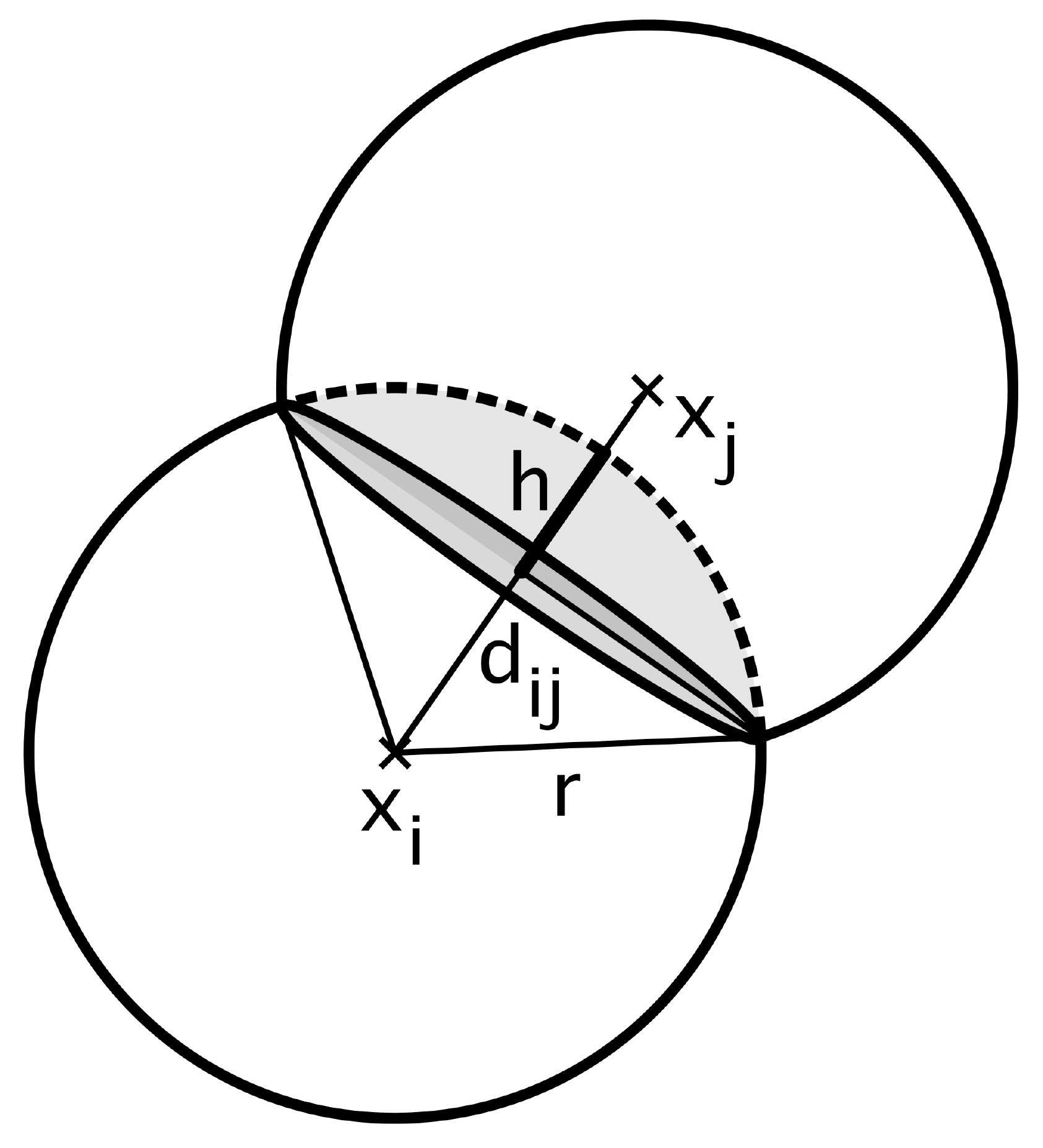} 
\includegraphics[width=0.53\linewidth]{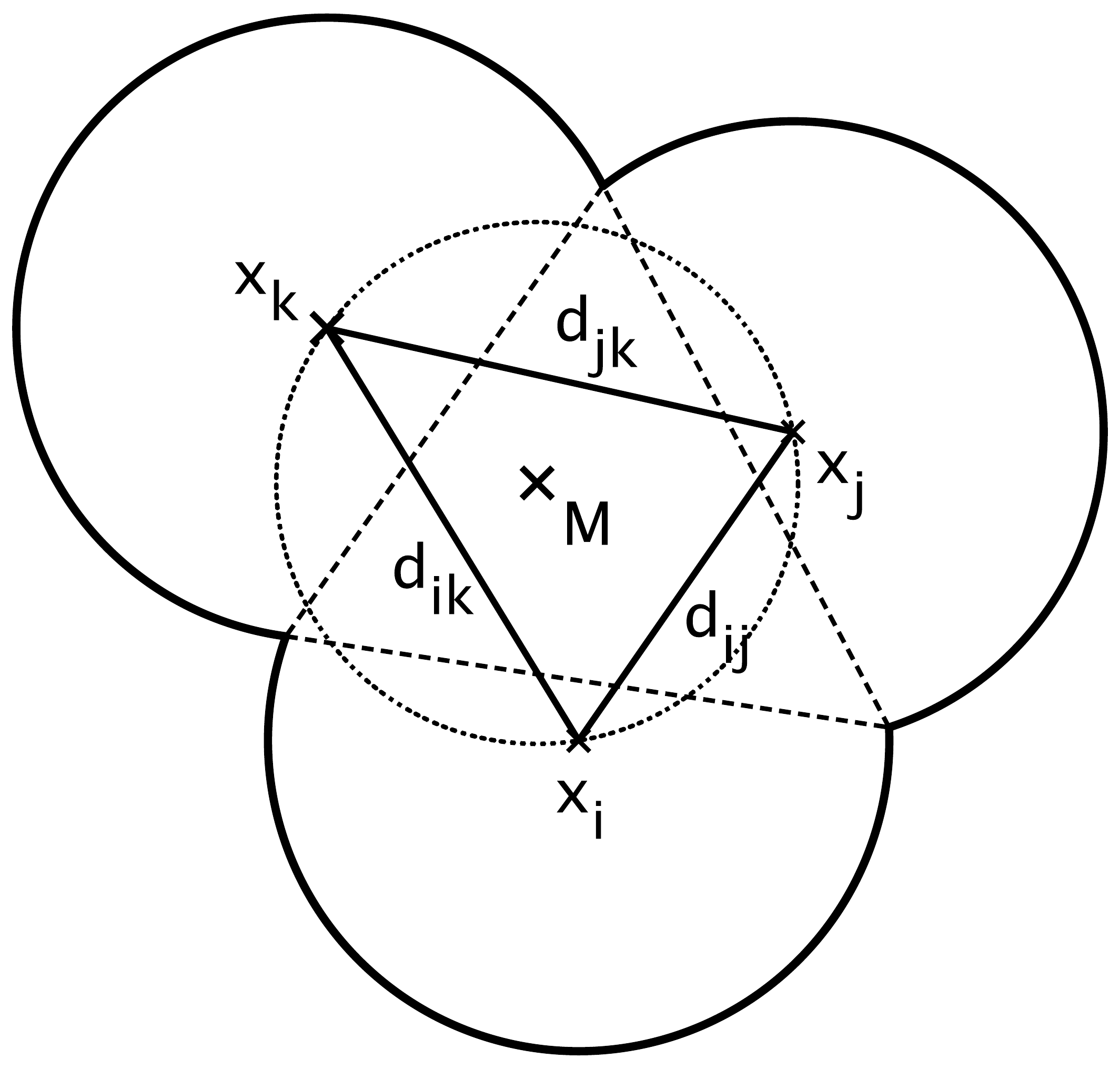}
\caption[Intersection of two resp.~three spheres]{The left--hand figure shows the covered surface area of a sphere
around $x_{i}$ when intersected with a second sphere as well as the
intersection circle. The right--hand figure illustrates the intersection
of three spheres. The triple point can be found above the centre point
of the circumcircle of the triangle generated by the three centre
points $\mathbf{x}_{i},\mathbf{x}_{j},\mathbf{x}_{k}$.}
\label{img:intersection} 
\end{figure}

The statistical weight of intersections of more than three Balls is
zero, see \cite{1994A&A...288..697M}; therefore, we only
take into account intersection circles of two Balls and intersection
points of three Balls, the so-called triple points, see Fig.~\ref{img:intersection}.
The MFs' volume densities are defined by 
\begin{equation}
v_{\mu}(\CB_{r})=\frac{1}{\mid\! D\!\mid}\sum_{i=1}^{N}V_{\mu}(\mathbf{x}_{i};r)\;,\label{mf-densities}
\end{equation}
 where $\mid\! D\!\mid$ denotes the volume of the sample mask.

In summary: for any given point $\mathbf{x}_{i}$ of the sample, we calculate
(for each radius $r$ up to the maximum radius $R$):
\begin{enumerate}
\item the uncovered surface area $A_{i}$ of the Ball around that point, 
\item the intersection circles of the Ball around that point with any of
its neighbours; here, $\ell_{ij}$ is the uncovered arc length, i.e.~the
uncovered segment of the intersection circle of the Balls around $\mathbf{x}_{i}$
and $\mathbf{x}_{j}$, 
\item the triple points of the intersection with the Balls around any two
neighbours, where $\epsilon_{ijk}$ is called spherical excess; it
can be calculated using the formula of l'Huilier and denotes the
contribution of the triple points to the Partial Euler characteriztic.
\end{enumerate}
With these quantities, the Partial MFs read (see \cite{1994A&A...288..697M,mecke2000statistical}
for more details):
\begin{eqnarray}
V_{1}(\mathbf{x}_{i};r) & = & \frac{A_{i}}{6}\;\;;\\
V_{2}(\mathbf{x}_{i};r) & = & \frac{A_{i}}{3\pi r}-\frac{1}{2}\sum_{j}\frac{d_{ij}\ell_{ij}}{6\pi}\;\;;\nonumber \\
V_{3}(\mathbf{x}_{i};r) & = & \frac{A_{i}}{4\pi r^{2}}-\frac{1}{2}\sum_{j}\frac{d_{ij}\ell_{ij}}{4\pi r\cdot\rho_{ij}}+\frac{1}{3}\sum_{j<k}\frac{\epsilon_{ijk}}{4\pi}\;\;,\nonumber 
\end{eqnarray}
where $d_{ij}=\; \parallel\mathbf{x}_{i}-\mathbf{x}_{j}\parallel\;\leq2r$
denotes the distance of two points, and $\rho_{ij}=\sqrt{r^{2}-(d_{ij}/2)^{2}}$
the radius of its intersection circle.

The use of the Partial MFs  also has the advantage that one can obtain an error estimate for the fluctuations of $V_{\mu}(\CB_{r})$ 
by calculating the variance of the values of the Partial MFs.

\begin{figure}
\centering 
\includegraphics[width=0.95\linewidth]{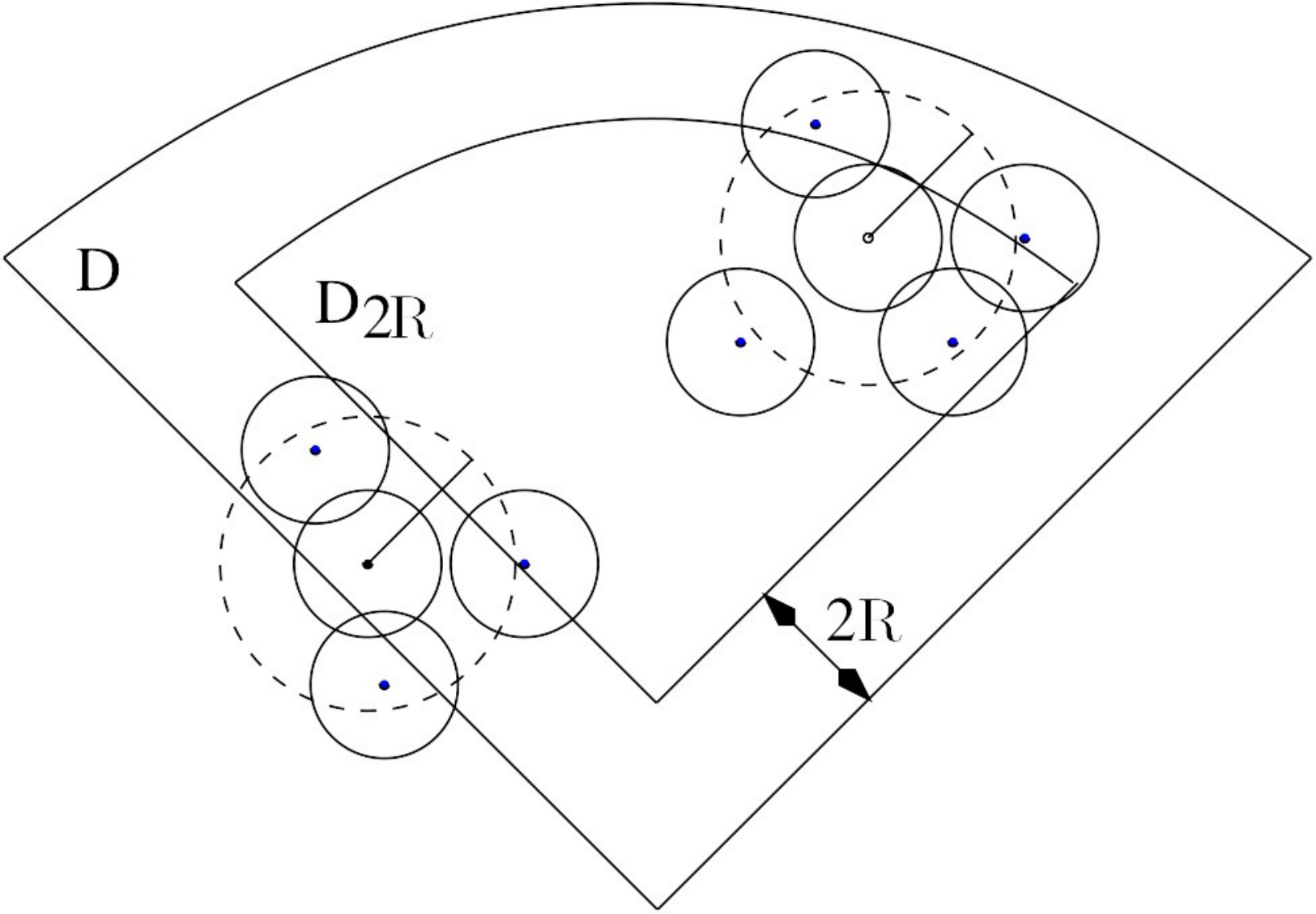}
\caption[Calculation of the Partial MFs]{To avoid boundary effects when calculating the Partial MFs, 
we only take those points into account, which are more
than two times the maximum Germ radius away from the survey mask.}
\label{img:partiell} 
\end{figure}

\subsection{Treatment of boundaries}

The family of MFs allows a complete deconvolution of the boundary, based on the principal kinematical formula, 
see for example the review of Kerscher~\cite[and references therein]{2000LNP...554...36K}. Note that this formula involves all
the functionals of the family; for individual functionals the boundary cannot be corrected with this powerful tool.
Most of the previously cited papers refer to this method for the boundary correction.
The principal kinematical formula reads: 
\begin{equation}
m_{\mu}(\CB_{r}) =  \frac{M_{\mu}(\CB_{r}\cap W)}{M_{0}(W)}-\sum_{\nu=0}^{\mu-1}{\mu \choose \nu}m_{\nu}(\CB_{r})\frac{M_{\mu-\nu}(W)}{M_{0}(W)}\;,
\end{equation}
where $M_{\mu}(\CB_{r})$ are the MFs, and $m_{\mu}(\CB_{r})$
their mean volume densities as defined in \cite{2000LNP...554...36K}.
$W$ denotes the boundaries, i.e.~the survey mask or window. For
an example illustrating these boundary corrections, we recommend \cite{1996ASPC...94..247K},
for an application to a galaxy catalogue see e.g.~\cite{1997MNRAS.284...73K}.

Unlike in these previous papers, we here calculate the Partial MFs only for points
more than \emph{twice} the maximum radius away from the boundary,
i.e.~the sample mask, see Fig.~\ref{img:partiell}. Thus, we create
a shrunk `calculation window' $D_{2R}$ and do not have to take
into account any boundary effects. Naturally, if the survey mask is
full of holes, we neglect a lot of galaxies this way, so this approach
is better suited for modern galaxy catalogues like the SDSS and after--SDSS surveys.

However, it is important to note that the neglected points do count
when it comes to calculating Partial MFs, since their Balls intersect
with Balls inside of the window. Therefore, they have to be part of the
neighbourlists.

The MFs volume densities~(\ref{mf-densities}) now take the form 
\begin{equation}
v_{\mu}=\frac{1}{\mid\! D_{2R}\!\mid}\sum_{i=1}^{N}\chi_{D_{2R}}(\mathbf{x}_{i})V_{\mu}(\mathbf{x}_{i};r)\;,
\end{equation}
 where 
\begin{equation}
\chi_{D_{2R}}(\mathbf{x}_{i})=\left\{ \begin{array}{ll}
1 & \textrm{if }\;\mathbf{x}_{i}\in D_{2R}\\
0 & \textrm{if }\;\mathbf{x}_{i}\notin D_{2R}
\end{array}\right.
\end{equation}
is the characteriztic function of the shrunk window. As mentioned
in \cite{2000MNRAS.312..638S}, these quantities are minus estimators
for the MF's volume-densities. Minus estimators
only provide unbiased estimates if applied to stationary point processes,
as investigated by \cite{1999A&A...343..333K}. Hence, we use volume-limited
subsamples of the catalogues when carrying out the structure analysis.

\subsection{The structure volume}

Since the volume of the structure is not localized on its surface,
we cannot calculate it in the way outlined above. However, to achieve analogy
to the three other functionals, and in view of the possibility of parallel
computing, our goal was to determine the volume by means of
adding up the partial functionals.

\begin{figure}
\centering 
\includegraphics[width=0.95\linewidth]{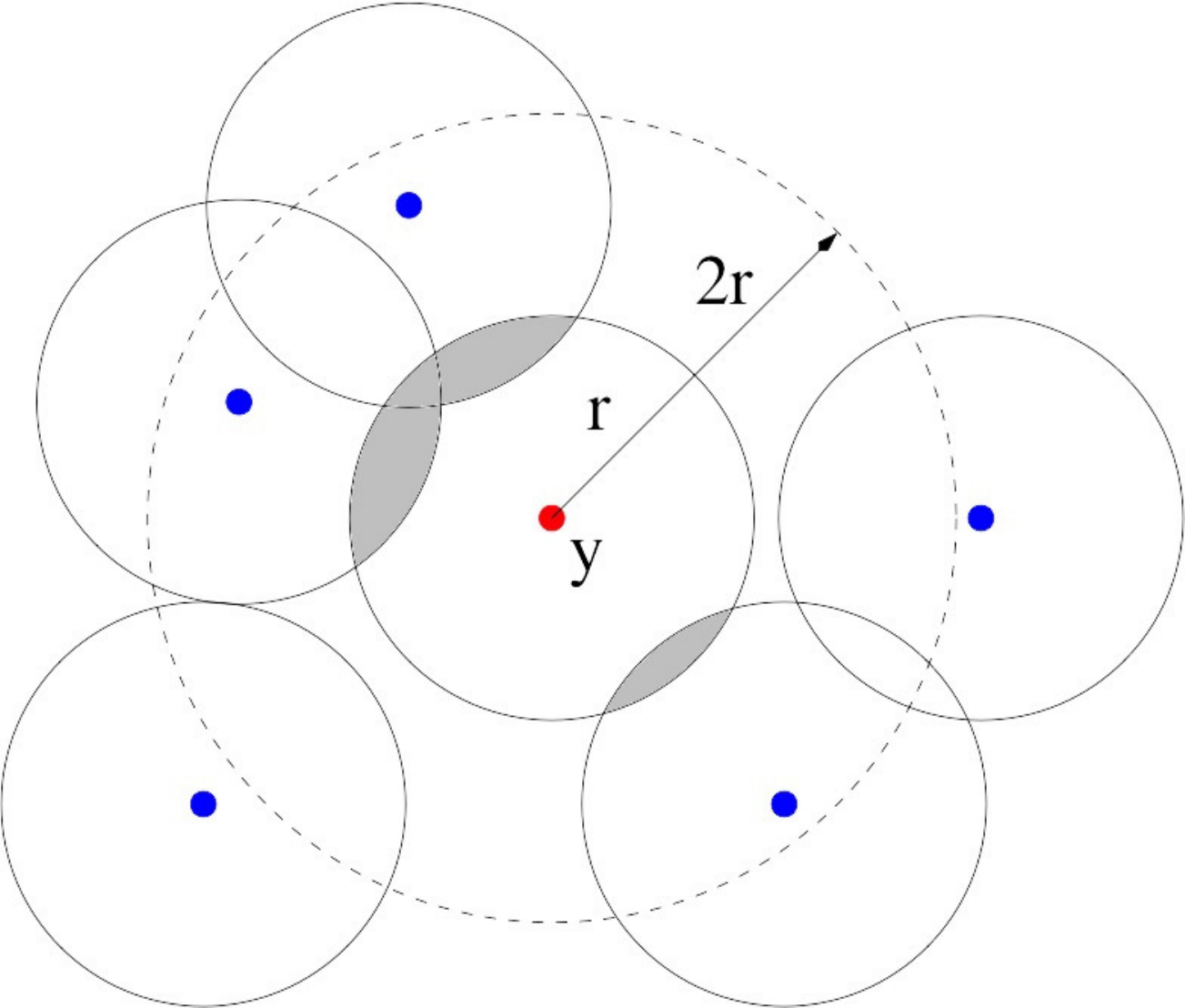}
\caption[Calculation of the structure volume.]{For any randomly thrown point $\textbf{y}_{i}$, we generate the
list of real neighbours, i.e.~consisting of galaxies of the
used sample. The Balls around all points within a distance of $2r$
around $\textbf{y}_{i}$ intersect with the Ball $\CB(\mathbf{y}_{i};r)$.
The volume fraction of the structure within $\CB(\mathbf{y}_{i};r)$
is then calculated by a Monte--Carlo integration.}
\label{img:v0} 
\end{figure}

We do this as follows: first, we throw a number of randomly distributed
points $\textbf{y}_{i}$ into the shrunk mask $D_{2R}$ of the sample;
secondly, we determine neighbourlists for the random points. These neighbourlists
consist of the thrown point as the centre and the \emph{real} galaxies
in its vicinity, i.e.~points of the sample we analyse, within a distance
of twice the maximum radius, see Fig.~\ref{img:v0}; thirdly, in a second Poisson process, we throw random points into the
Ball around $\mathbf{y}_{i}$, i.e.~$\CB(\mathbf{y}_{i};r)$, and
determine whether the random point is covered by a Ball around any
of the \emph{real} neighbours or not. This way we calculate the fraction
of volume covered by the structure in that local area. Hence, we defined
a Partial MF $v_{0}(\mathbf{y}_{i};r)$ for the volume similar to
the other three (strictly speaking, we defined the volume density
of the Partial MF); the last step for obtaining the global volume density of the structure
consists of adding up the $v_{0}(\mathbf{y}_{i};r)$ and normalizing
them by the number of random points, say $M$, for which we calculated
the volume fraction, 
\begin{equation}
v_{0}=\frac{1}{M}\sum_{i=1}^{M}v_{0}(\mathbf{y}_{i};r)\;.
\end{equation}
If we are interested in the absolute value of the structure's volume
within the shrunk windows, we obtain it by multiplying the global
volume fraction by the window's volume.
Note: only the third and fourth step of the volume fraction calculation
are executed by the \textsc{\footnotesize CHIPMINK} code itself, whereas the primary steps are
data preparation. So instead of throwing points into the shrunk window
$D_{2R}$ in the first step, one can throw them into the original
survey mask and create neighbourlists for all of them. Thus, calculations
for different maximum radii or specific areas of the survey mask can
be carried out with subsamples of this set of neighbourlists.

For consistency checks, we also provide two other methods for the calculation of the structure volume. However, these methods only calculate the global volume fraction and they do not use the Partial MFs defined in this section.

\section{Application: the SDSS LRG sample}
\label{sec:Minkowski-fun-LRG}
 
We shall now apply the code described in the previous section to two
different data sets. First, to the luminous red galaxy (LRG, \cite{2001AJ....122.2267E}) sample
of the Sloan Digital Sky Survey (SDSS, \cite{2000AJ....120.1579Y}) Data release 7 (DR7, \cite{2009ApJS..182..543A}),
and, secondly, to the mock catalogues drawn from $\Lambda$CDM simulations
of the SDSS volume performed by the LasDamas\footnote{See \href{http://lss.phy.vanderbilt.edu/lasdamas/}{http://lss.phy.vanderbilt.edu/lasdamas/} for information on the project and for downloading the samples.} collaboration \cite{lasdamas}.

\subsection{The data}

From the SDSS DR7 LRG data described in \citet{2009ApJS..182..543A},
we use in particular the samples extracted by \citet{2010ApJ...710.1444K}.
For selecting them the authors used the following criteria:
the galaxy has an SDSS spectrum, is not in an area around bright stars,
has a sector completeness of at least 60\%, a redshift in the range
$0.16-0.47$ and a colour-- and $k$--corrected magnitude between $-21.2$
and $-23.1$. The details of the selection can be found in \cite{2010ApJ...710.1444K}.
After this pre--selection, the sample contains $105\,831$ LRGs.

We neglect the small amount of area in the southern galactic regions from the DR7 sample, as our code requires contiguous regions
(note that boundaries can be corrected by
integral--geometrical means; this property is exploited in our previous
codes). In addition, in order to have simpler boundaries we choose to restrict
ourselves to a region with $\text{RA}\in\left[132^{\circ},235^{\circ}\right]$
and $\text{Dec.}\in\left[-1^{\circ},60^{\circ}\right]$. 

With the radial selection we have to make sure that the sample we
obtain is volume limited. According to \cite{2010ApJ...710.1444K},
the sample is volume limited up to a redshift of $0.36$ for a magnitude
of $-21.2$ and up to a redshift of $0.44$, if we select galaxies
with a magnitude brighter than $-21.8$. This implies that we shall analyse
two different samples derived from the pre--selected galaxies: a first
one that we shall refer to as the `dim sample' with a magnitude
cut at $-21.2$ and a redshift in the range $z\in\left[0.16,0.35\right]$,
and a second one to which we refer to as the `bright sample' with
a magnitude cut at $-21.8$ and a redshift in the range $z\in\left[0.16,0.44\right]$.
With these requirements, the `dim sample' contains $41\,375$ galaxies
and the `bright sample' $22\,386$. 

In order to compare the structure in the galaxy data to the model
of gravitational structure formation, we use mock data samples provided
by the LasDamas\footnotemark[\value{footnote}]
collaboration \citep{lasdamas}.  
These authors simulated structure formation in a $\Lambda$CDM model
with $\Omega_{\Lambda}=0.75$ in boxes of $2.4\hGpc$ for $1280^{3}$
particles. 
They identify haloes with a friends--of--friends algorithm,
and populate them with mock galaxies using a halo occupation distribution (HOD). 
The HOD--parameters are chosen as to reproduce small-scale clustering of the observed LRG sample.
From 40 independent $N$--body simulations, LasDamas provides 160 sky--based mock galaxy catalogues for the northern SDSS region that we use.

\begin{figure}
\includegraphics[width=0.45\textwidth]{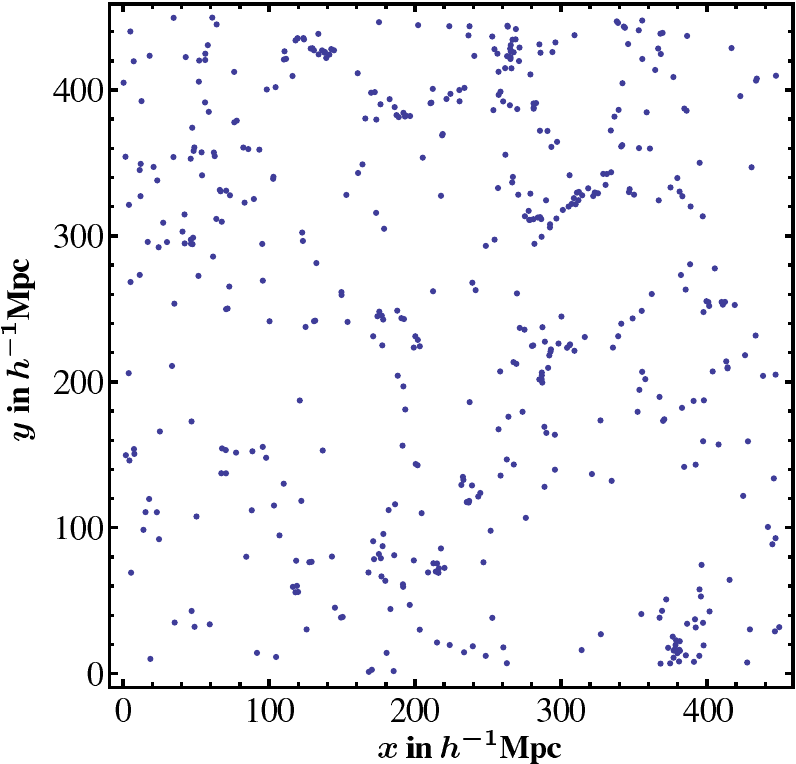}

\caption{Example of the galaxy distribution in the `dim sample'. The projection is extracted from a slice of a thickness of $22.5\hMpc$ from the maximally fitting cube.
\label{fig:Galaxies}}
\end{figure}

We further modify the basic catalogues they provide by also removing regions around bright stars\footnote{We use the software mangle \cite{2008MNRAS.387.1391S} to apply the
mask that can be found in the NYU value--added catalogue \cite{2005AJ....129.2562B,2008ApJ...674.1217P}.} 
and performing the same angular cut as for the SDSS data samples. 
This results in catalogues that contain on average $46\,710$ galaxies for the `dim sample' and $22\,181$ galaxies for the `bright sample'.

Our treatment of various issues encountered in data taking, which should be included in the mock galaxy selection, is not complete. It does not take problems like e.g. sector completeness or fibre collisions into account. To deal with these issues the necessary weights could probably be included into Equation~(\ref{eq:sumpart}), but we leave it for a more thorough analysis in future work to figure out the exact form of this weighting.
For the purpose of this paper, testing the code and a general analysis of the influence of higher clustering we do not need this precision.

Instead of considering the MFs in redshift space,
we convert all redshifts into comoving distances using the distance
redshift relation of a $\Lambda$CDM model with $\Omega_{\Lambda}=0.75$.
An example of the galaxy distribution in the `dim sample' is shown in Fig.~\ref{fig:Galaxies}. It also
helps to recall the dimensions of the sample. The thickness of the
$z$--shell of the \textquotedbl{}dim sample\textquotedbl{} $z\in\left[0.16,0.35\right]$
is $507\hMpc$. The thickness of $z$--shell of the \textquotedbl{}bright
sample\textquotedbl{} $730\hMpc$. The largest cube that fits into
our \textquotedbl{}dim sample\textquotedbl{} region has a side length
of $452\hMpc$; Fig.~\ref{fig:Galaxies} presents a slice of this cube. 
In \cite{wiegandphd} two independent cubes of this size have been used to demonstrate the stability of the MFs throughout the sample.

\subsection{The functionals on different scales\label{sub:The-functionals}}

We now turn to the analysis of the samples defined in the previous
section. In this analysis we compare the structure in the observed
samples to the structure in the mock samples. For this comparison
it is crucial to know, how precise our results for the MF
densities are. To estimate these errors we determine the MF densities 
for each of the 160 mock samples and calculate the error bars from 
the resulting fluctuation.
For comparison, we also calculated the error bars from random subsampling {\em jackknife}
realizations drawn from the data and consisting of 80\% of the points of the samples.
They turn out to be of the same magnitude.
Finally, we also compared them to the error estimate that the \textsc{\footnotesize CHIPMINK} code determines
directly from the fluctuations of the Partial MFs. Also in this case, the error bars are close 
to those determined from the mocks, even though systematically smaller by a few percent.
So, for a first estimate of the errors already the output of the code is quite useful.

There are two possible reasons for the MF densities
(\ref{eq:MinkDensDef}) to fluctuate between different realizations.
First, the $N$--point correlation functions of the point distribution 
in different realizations may be different.
Then, the integrals~(\ref{eq:MinkIntegr}) and therefore the coefficients
in the expansion~(\ref{eq:MinkCorrCon}) vary and lead to fluctuations
in the measured $v_{\mu}$. But the series~(\ref{eq:MinkCorrCon})
indicates that also a different average density $\varrho_{0}$ of
the sample will lead to a change of the measured $v_{\mu}$. This
means that we have to ensure that all the realizations approximately
have the same density, if we really want to compare the structure
of the point distribution. Otherwise, the analysis of the influence
of (higher order) correlations in the point distribution would be
spoiled by a fluctuation in $\varrho_{0}$.

To ensure that all realizations have the same density $\varrho_{0}$,
we implement a random choice of $\approx80\%$ of the sample that discards 
all configurations that do not have the desired density. Due to slightly different
average densities of the mock and data samples this fraction is not exactly $80\%$,
but is adjusted to give the same average density for the mocks and the data.

\subsubsection{The `dim sample'}

\begin{figure*}
\includegraphics[width=0.85\textwidth]{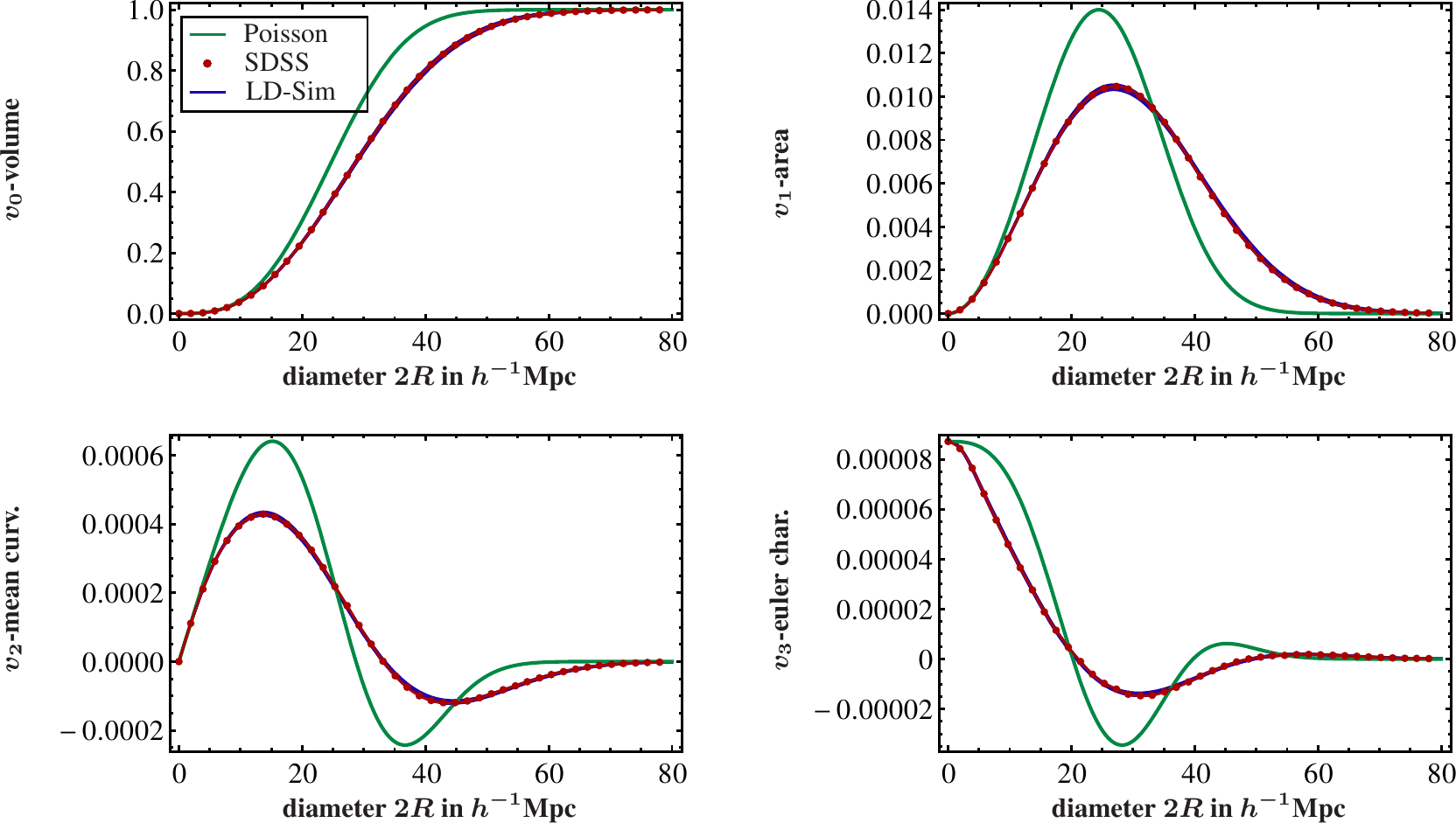}
\includegraphics[width=0.85\textwidth]{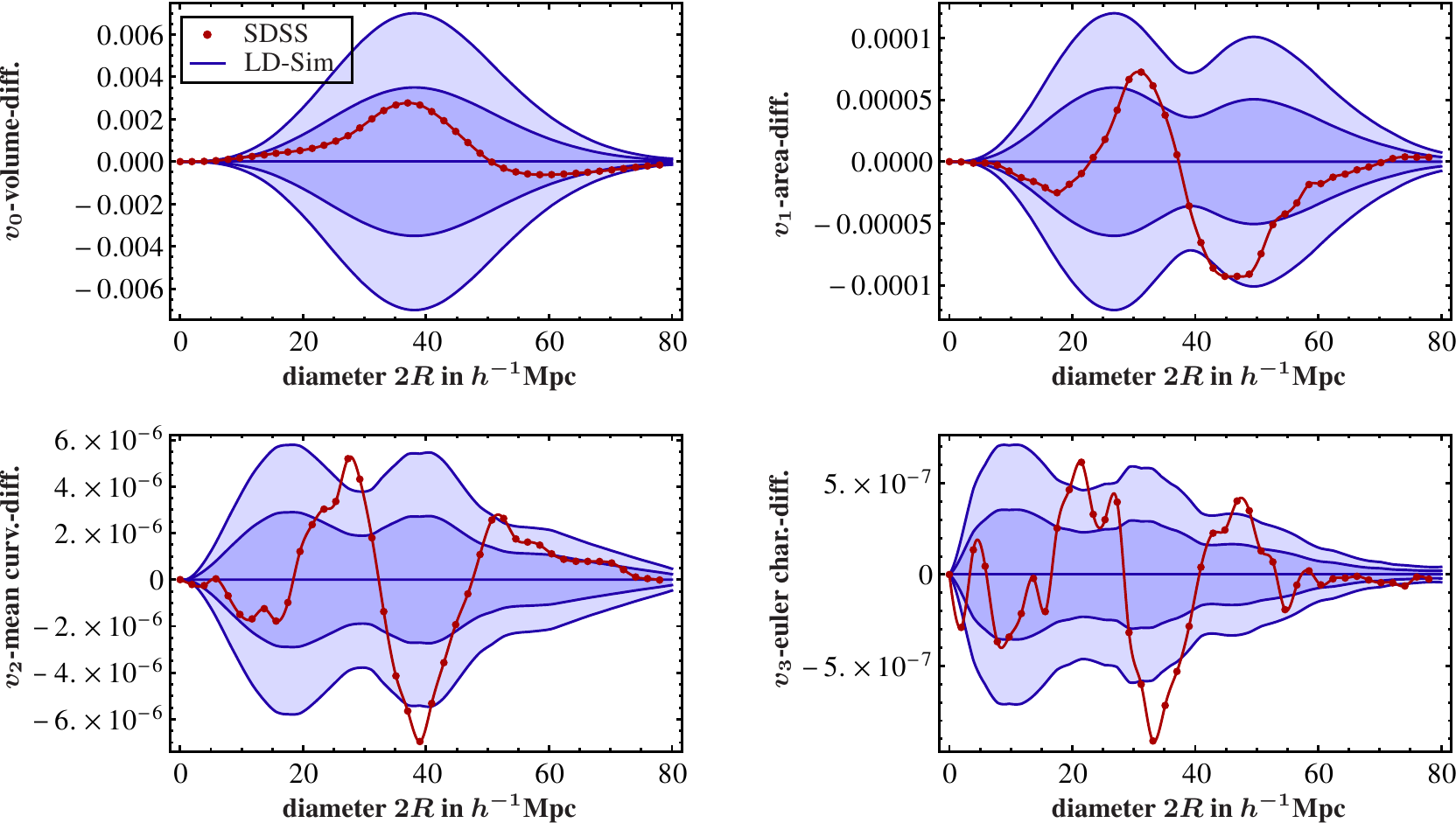}
\caption{Top four: the four Germ--Grain MF densities for the
SDSS LRG `dim sample' compared to those of the corresponding LasDamas
mock galaxies and those of a Poisson distribution. The errors and average for the mock samples are
obtained taking $160$ different mock realizations.\protect \\
Bottom four: the same quantities, but with the average of the mock sample
subtracted to make the error bars more visible. The dark shaded regions
are the $1\sigma$, the light shaded regions $2\sigma$ error bands.\label{fig:full-av}}
\end{figure*}

Fig.~\ref{fig:full-av} shows
the MF densities
obtained from this procedure and the code described in Section~\ref{sec:code}.
\begin{figure*}
\includegraphics[width=0.85\textwidth]{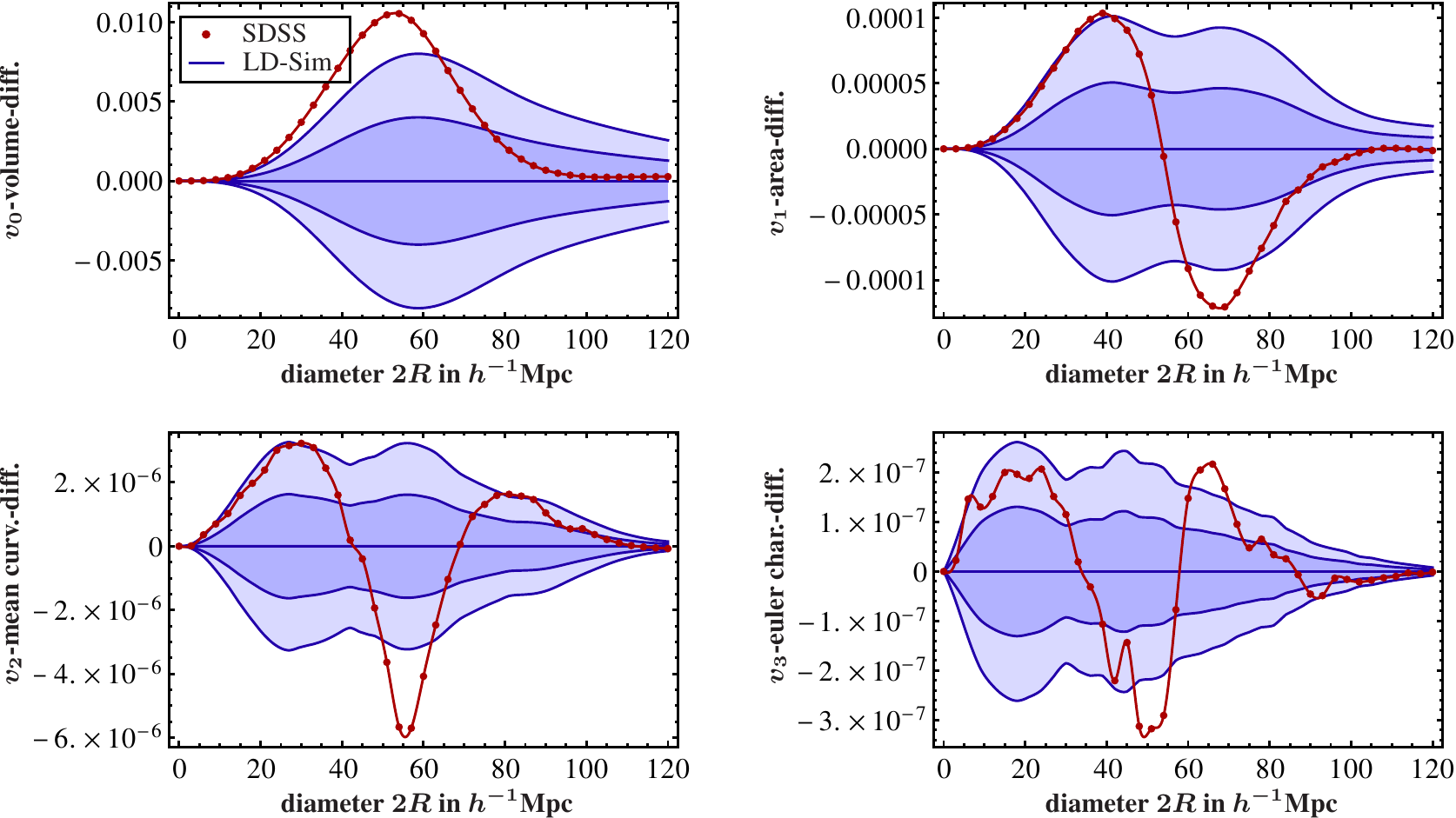}
\caption{The same quantities as in the lower four plots of Fig.~\ref{fig:full-av},
but for the bright ``sample''. The dark shaded regions
are the $1\sigma$, the light shaded regions $2\sigma$ error bands.\label{fig:bright-av}}
\end{figure*}
We plot the average from $244$ random $80\%$--realizations of the `dim sample'. The functionals were evaluated for Balls of $40$ different radii with a spacing of about $1\hMpc$. They are shown as the (red) points in Fig.~\ref{fig:full-av}. 

The selection of 80\% of the points results in an average density of $\varrho_{0}=7.7\times10^{-5}\hDens$.
For the mock catalogues, we also extract $244$ configurations of this average density $\varrho_{0}$
from each of the $160$ simulated samples and calculate the average value for the MFs of each mock with the same radii of the Balls.
The mean value and the error bars are then calculated from the mean and the variance of these $160$ averages.

The upper four plots in Fig.~\ref{fig:full-av} indicate that the
determination of the MF densities is quite robust.
The $1\sigma$ error bands around the average are barely visible.
The values for a Poisson distribution with the same density lie far
away from both mocks and observed galaxies, a clear indication of
the presence of structure (which is of course not surprising). For
the curves of the mocks and observed galaxies, however, it is harder
to distinguish them.

In order to facilitate the comparison of the observed and mock results, the
lower four plots of Fig.~\ref{fig:full-av} show the residuals obtained by subtracting the average of the mocks. Most of the $40$ points lie in the $1-$ and $2-$$\sigma$ bands around the average of the mocks, but some points deviate more strongly. 

In order to quantify how significant the resulting deviation actually is, we calculate the $\chi^2$ values using the standard relation,
\begin{equation}
\chi^2 :=\sum_{i j}\left(v^d_\mu(r_i)-\ol{v}^m_\mu(r_i)\right) \hat{C}_{i j}^{-1} \left(v^d_\mu(r_j)-\ol{v}^m_\mu(r_j)\right),
\label{eq:chisq}
\end{equation}
where $v^d_\mu(r_i)$ are the functionals measured from the SDSS, $\ol{v}^m_\mu(r_i)$ is the average of all mocks and $\hat{C}_{i j}^{-1}$ is the inverse covariance matrix of the $40$ points, estimated from the $160$ mock realizations. We use the unbiased estimator of \cite{2007A&A...464..399H} for $\hat{C}^{-1}$, which takes into account that the estimate is based on a finite number of mock samples. 

The resulting $\chi^2$ values are shown in the first column of Table~\ref{tab:dimvsigma}. They are significantly higher than the average of the $\chi^2$ distribution of $40$ points. In order to quantify the deviation, we convert the $p$-value corresponding to the value of $\chi^2$ into standard deviations of a Gaussian random variable, where the $p$-value is taken to correspond to the two sided deviation. The deviation is larger than $3\sigma$ for all of the four MFs. As this is surprising regarding the plots, and to illustrate the amount of correlation in the data, we also list the $\chi^2$ values we would obtain, if the data points were independent. We do this by taking the diagonal of the covariance matrix before inverting it. The resulting $\chi^2_I$ values are in good agreement with the visual impression given the residuals in Fig.~\ref{fig:full-av}.

\begin{table}
\center
\begin{tabular}{p{0.02\linewidth}p{0.04\linewidth}p{0.04\linewidth}|p{0.04\linewidth}p{0.04\linewidth}p{0.02\linewidth}p{0.04\linewidth}p{0.04\linewidth}p{0.04\linewidth}|p{0.04\linewidth}p{0.05\linewidth}}
\cline{1-5} \cline{7-11} 		\noalign{\smallskip} 		

 & $\chi^2$ & $\sigma_G$ & $\chi^2_I$ & $\sigma_{G\,I}$& & & $\chi^2$ & $\sigma_G$ & $\chi^2_I$ & $\sigma_{G\,I}$\\
\noalign{\smallskip} 		\cline{1-5} \cline{7-11} 		\noalign{\smallskip}

 $v_0$ & 78.7 & 3.66 & 21 & $10^{-2}$ & & $\ol{V}_0$ & 43.9 & 3.17 & 11 & 0.1 \\
 $v_1$ & 111. & 5.65 & 33 & 0.3 & & $\ol{V}_1$ & 68.7 & 5.13 & 15 & 0.2 \\
 $v_2$ & 81.2 & 3.83 & 53 & 1.7 & & $\ol{V}_2$ & 65.1 & 4.87 & 17 & 0.4 \\
 $v_3$ & 93.3 & 4.62 & 68 & 2.9 & & $\ol{V}_3$ & 62.9 & 4.71 & 30 & 1.8 \\

\noalign{\smallskip} 	\cline{1-5} \cline{7-11} 	\noalign{\smallskip}

\end{tabular}%
\caption{$\chi^2$ values for the deviation of the `dim sample' from the LasDamas mocks for the MF densities $v_\mu$ and the modified functionals $\ol{V}_\mu$. $\sigma_G$ quantifies the $p-$value that corresponds to this $\chi^2$, in terms of standard deviations of a Gaussian. $\chi^2_I$ quantifies the deviation, if we would assume statistical independence of the points. The data for $v_\mu$ are plotted in the lower panel of Fig.~\ref{fig:full-av}. The $\chi^2$ is done over $40$ degrees of freedom for $v_\mu$, and over $20$ degrees of freedom for $\ol{V}_\mu$.}	\label{tab:dimvsigma} 
\end{table}

\subsubsection{The `bright sample'}

We applied the same procedure to the `bright sample'. In this case,
randomly choosing 80\% of the points corresponds to a density of $\varrho_{0}=2.1\times10^{-5}\hDens$.
The resulting differences of the MF densities of
the mocks to those of the SDSS `bright sample' are shown in Fig.~\ref{fig:bright-av}.
The reduced galaxy density allows us to go to larger scales, due to the
larger volume and a more restrictive selection. This is possible,
because it needs a larger radius of the Balls to fill the observed
volume completely. Therefore, the structure saturates for larger values
of $R$ only. The downside is, however, that we have less points and therefore a less precise determination of the average functionals.
The plots in Fig.~\ref{fig:bright-av} show a similar deviation as those for the `dim sample'. When we evaluate the significance of this discrepancy, however, we find that it is less pronounced. The $\chi^2$ values calculated with equation~(\ref{eq:chisq}) and shown in Table~\ref{tab:brightvsigma}, are much lower than for the `dim sample'. This is somewhat surprising but, apparently, the deviations follow more closely the form imposed by the correlation pattern.

\begin{figure*}
\includegraphics[width=0.85\textwidth]{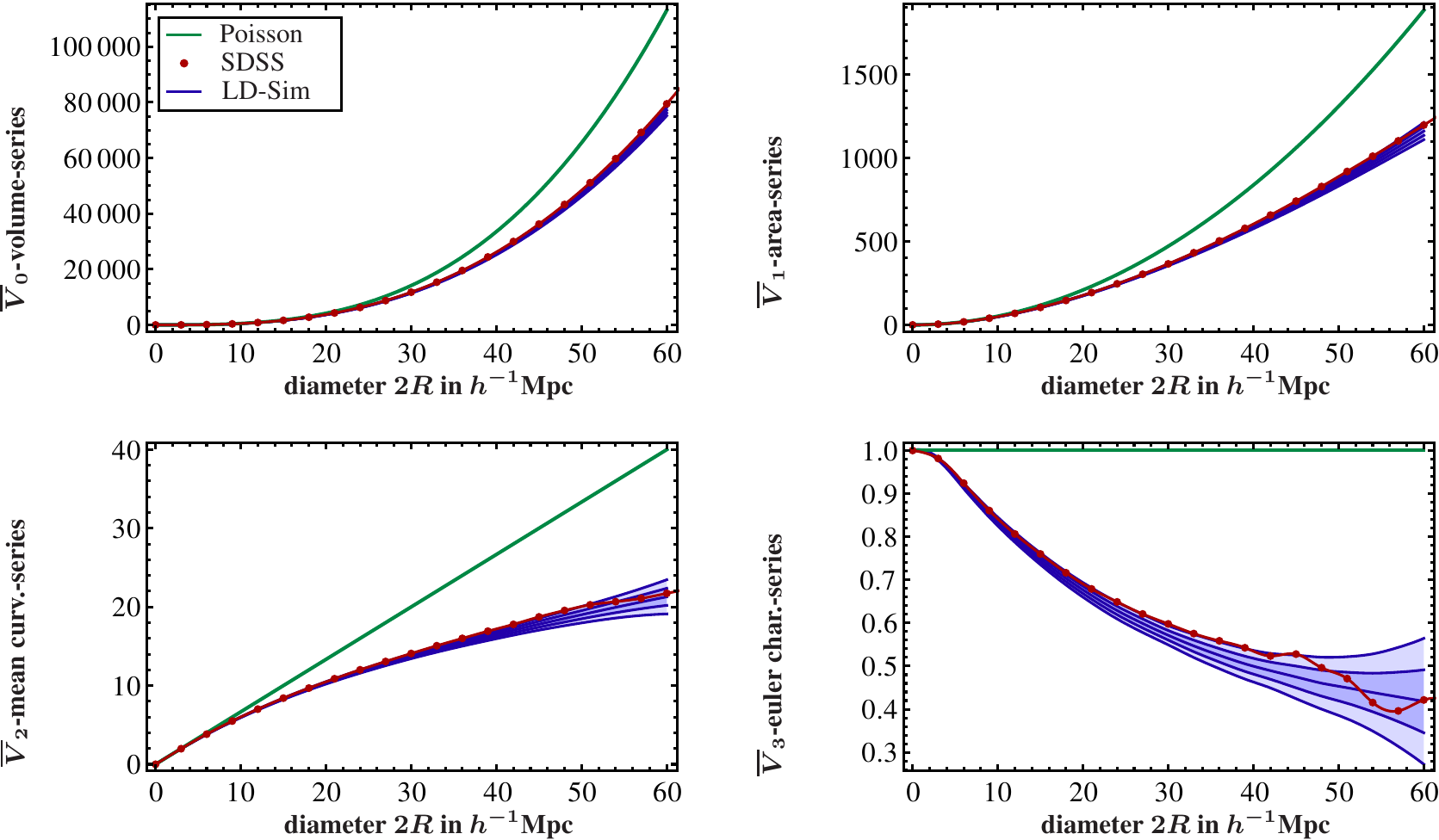}
\includegraphics[width=0.85\textwidth]{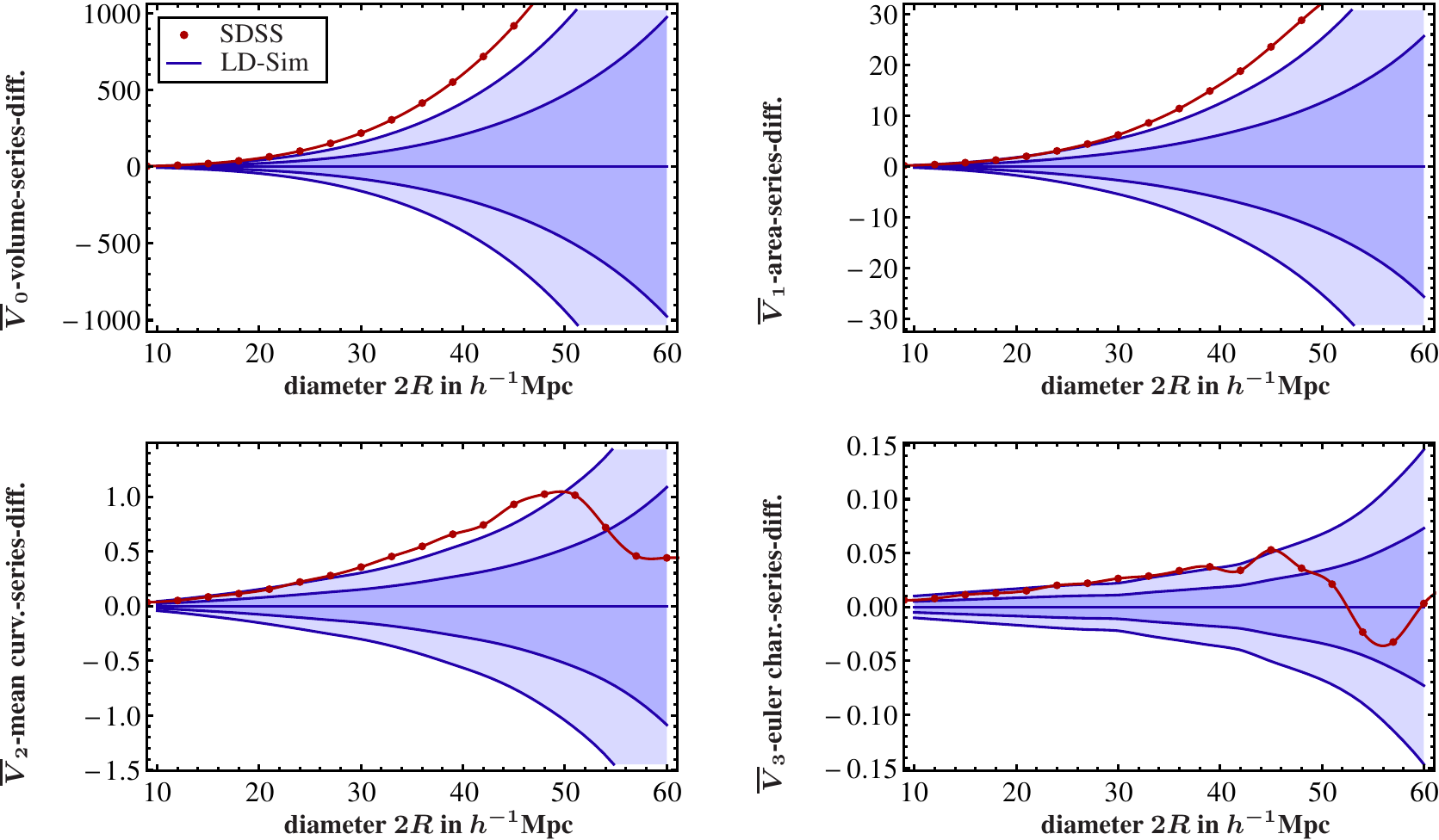}
\caption{Top four: the four modified Germ--Grain MFs $\overline{V}_{\mu}$
of the SDSS LRG `bright sample' and the LasDamas mock galaxies.
The errors and average for the mock samples are obtained taking $160$ different mock realizations.
The Poisson case is simply given by the MFs of a single ball (see Equation~(\ref{eq:MinkBall})). \protect \\
Bottom four: the same quantities, but with the average of the mocks subtracted to make the error bars more visible.
The dark shaded regions are the $1\sigma$ error bands, the light
regions correspond to $2\sigma$.\label{fig:bright-min}}
\end{figure*}%

\begin{table}
\center
\begin{tabular}{p{0.02\linewidth}p{0.04\linewidth}p{0.04\linewidth}|p{0.04\linewidth}p{0.04\linewidth}p{0.02\linewidth}p{0.04\linewidth}p{0.04\linewidth}p{0.04\linewidth}|p{0.04\linewidth}p{0.05\linewidth}}
\cline{1-5} \cline{7-11} 		\noalign{\smallskip} 		

 & $\chi^2$ & $\sigma_G$ & $\chi^2_I$ & $\sigma_{G\,I}$& & & $\chi^2$ & $\sigma_G$ & $\chi^2_I$ & $\sigma_{G\,I}$\\
\noalign{\smallskip} 		\cline{1-5} \cline{7-11} 		\noalign{\smallskip}

 $v_0$ & 56.6 & 2.02 & 105 & 5.4 & & $\ol{V}_0$ & 35.3 & 2.36 & 112 & 7.8 \\
 $v_1$ & 52.6 & 1.71 & 74 & 3.3 & & $\ol{V}_1$ & 26.0 & 1.38 & 80 & 5.9 \\
 $v_2$ & 50.8 & 1.56 & 82 & 3.9 & & $\ol{V}_2$ & 31.7 & 1.99 & 66 & 4.9 \\
 $v_3$ & 36.0 & 0.46 & 70 & 3.0 & & $\ol{V}_3$ & 28.4 & 1.65 & 48 & 3.5 \\

\noalign{\smallskip} 	\cline{1-5} \cline{7-11} 	\noalign{\smallskip}

\end{tabular}

\caption{Same as Table~\ref{tab:dimvsigma}, but for the `bright sample'. The data for $v_\mu$ are plotted in Fig.~\ref{fig:bright-av} and those for $\ol{V}_\mu$ in the lower panel of Fig.~\ref{fig:bright-min}. The $\chi^2$ is done over $40$ degrees of freedom for $v_\mu$, and over $20$ degrees of freedom for $\ol{V}_\mu$.}	\label{tab:brightvsigma} 
\end{table}

As this correlation pattern is different for the modified MFs $\overline{V}_{\mu}$, we also want to study their deviations from the mock samples and show them in Fig.~\ref{fig:bright-min}.
The upper four plots show them together with the Poisson case. This
latter is very simple for $\overline{V}_{\mu}$ as it consists
of the functionals of a Ball, Equation~(\ref{eq:MinkBall}). The values
and error bars have been obtained by calculating the $\overline{V}_{\mu}$
for every realization and taking the average and variance of these
values. As the errors grow rapidly beyond $60\hMpc$ (in diameter), we only plot
the $\overline{V}_{\mu}$ up to this scale. The reason for this
growth is that around $60\hMpc$, the volume becomes largely filled
with the Balls, and therefore, the measurement has to become more and
more accurate to give correct values after the removal of the exponential
damping factor $e^{-\varrho_{0}\overline{V}_{\mu}}$. 
Considering the scales where the errors are still controllable leaves us with the $20$ points in Fig.~\ref{fig:bright-min}. The $\chi^2$ values of their deviation are summarized in the second part of Table~\ref{tab:brightvsigma}. The reduced $\chi^2$ values are a bit higher than for the densities $v_\mu$, but also not significant.

\section{Non--Gaussian correlations}
\label{sec:Non-Gaussian-correlations}

Inspecting Figures~\ref{fig:bright-av} and~\ref{fig:bright-min}, it is not clear whether the origin of the slight deviation of the data from the mocks is already present for the two--point statistics, or whether it is due to a difference in the non--Gaussian properties showing up in the three-- and $N-$point contributions to the series~(\ref{eq:MinkCorrCon}).
To address this question more thoroughly, we use the method introduced in Section~\ref{sub:Extracting-higher-order}.
This method requires to measure the MFs as a function
of $\varrho_{0}$, and to extract for every scale $R$ the function
$\tilde{\overline{V}}_{\mu}\left(\varrho_{0}\right)$.

The procedure we use is the following: first, we choose $24$ different
densities corresponding to a fraction $f$ between $0.05$ and $0.8$
of the full density of $\varrho_{0}=2.6\times10^{-5}\hDens$. For
each of these densities, we generate a large number of realizations
(from about $15000$ for $f=0.05$ down to $244$ for $f=0.8$).
For each of these realizations we determine the modified MFs
$\overline{V}_{\mu}$ as a function of the Ball radius
$R$. For each $R$ we take the average over all the realizations
and arrive at an $\tilde{\overline{V}}_{\mu}\left(\varrho_{0}\right)$
evaluated at $24$ points. We add the value of $\ol V_{\mu}$ at $\varrho_{0}=0$
as the $25$th point, which is simply the MF of
a Ball (see Equation~(\ref{eq:MinkBall})) with the respective radius
$R$. From these $25$ points we want to derive the first coefficients
in the expansion~(\ref{eq:MinkCorrCon}) which, by Equation~(\ref{eq:MinkIntegr}),
is equivalent to the determination of the components of a polynomial
fit to $\tilde{\overline{V}}_{\mu}\left(\varrho_{0}\right)$. %
\begin{figure*}
\includegraphics[width=0.9\textwidth]{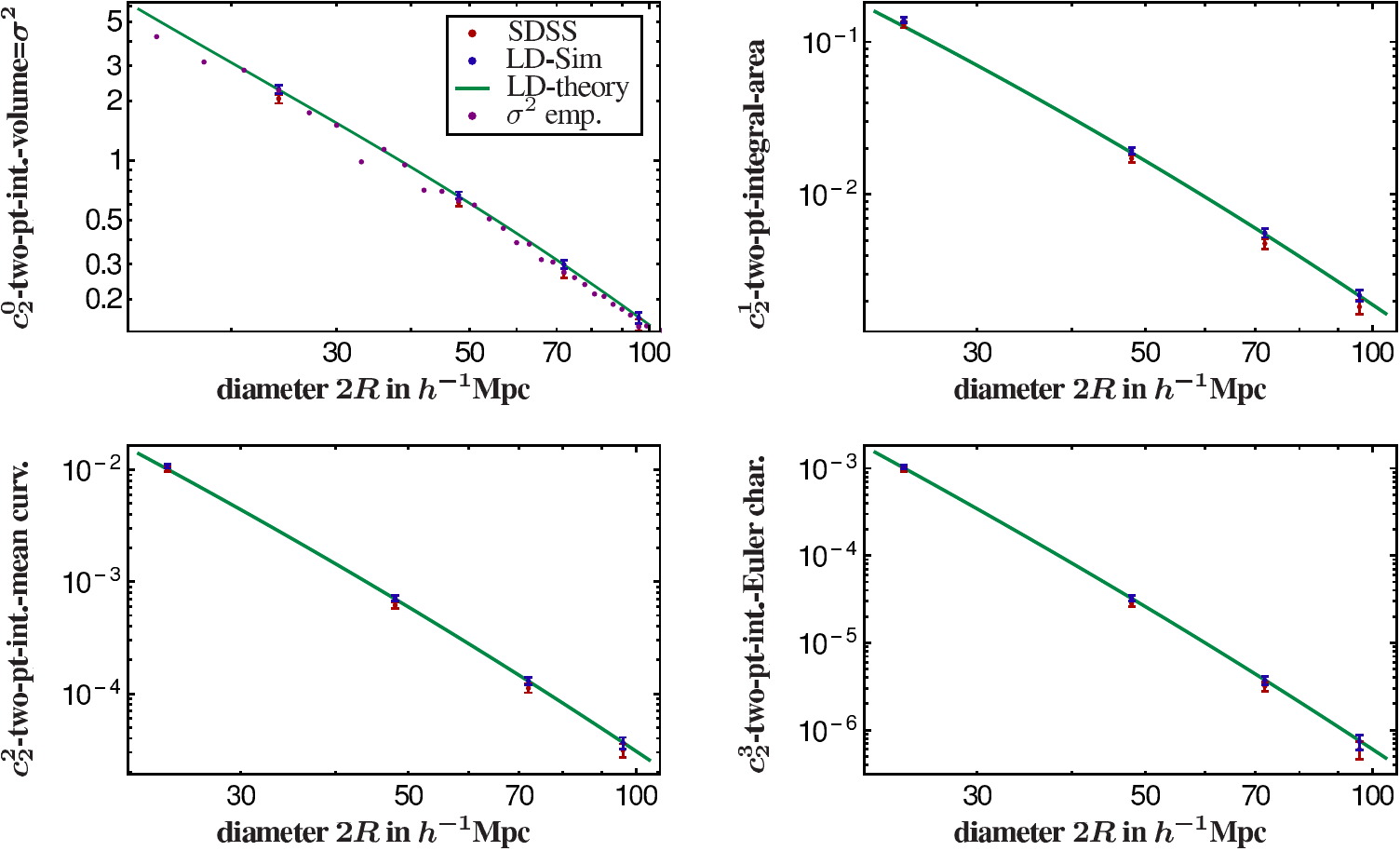}
\includegraphics[width=0.9\textwidth]{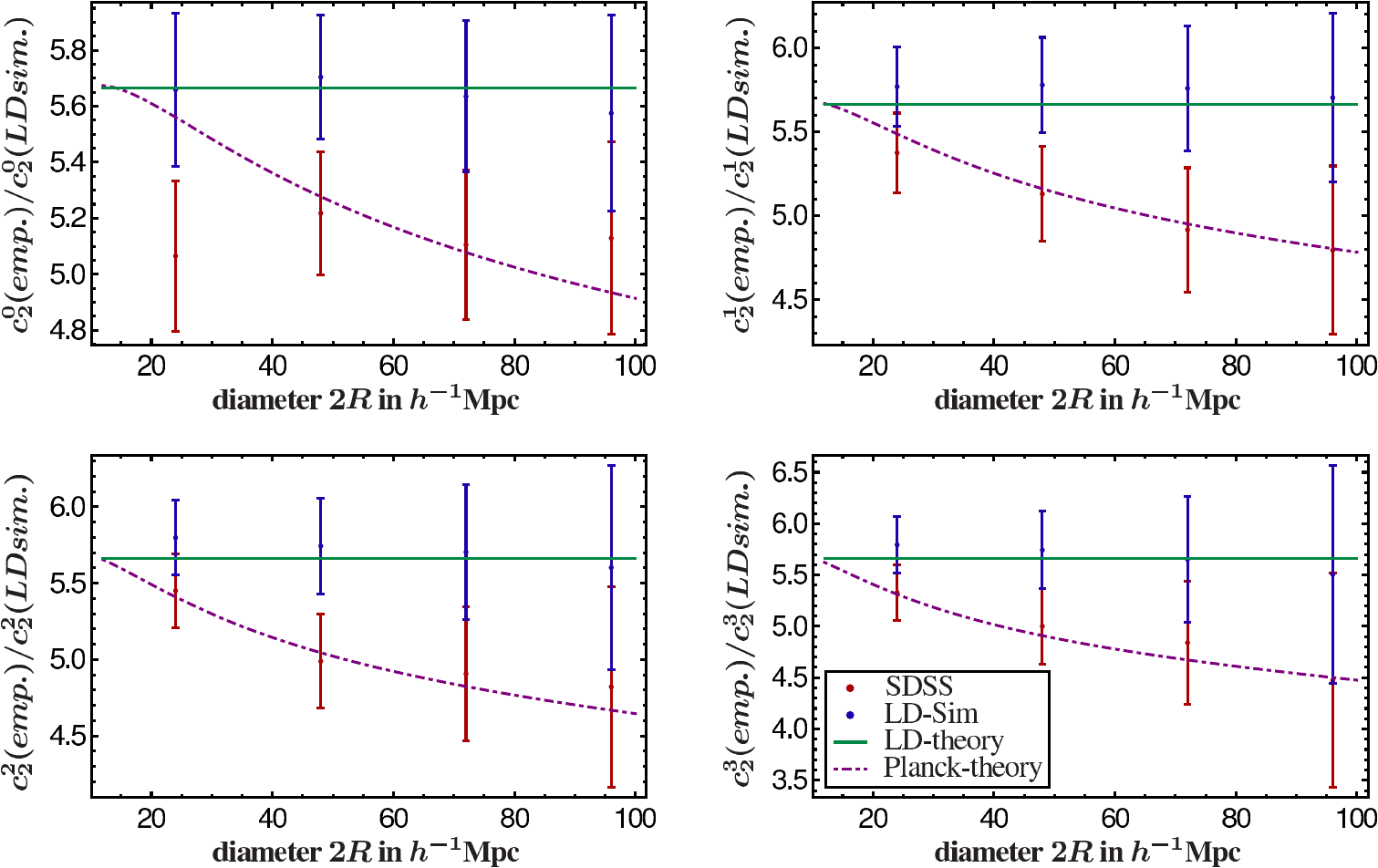}
\caption{
Top four: the four Minkowski integrals of the two--point correlation
function, i.e. $c_{2}^{\,\mu}$ from Equation~(\ref{eq:NormInteg}),
for the SDSS LRG `bright sample' (lower points, red) and one of the corresponding
mock samples (upper points, blue). The error bars are $1\sigma$ errors for the parameters
from the fit $\tilde{\overline{V}}_{\mu}\left(\varrho_{0}\right)$
with a polynomial of fourth order in $\varrho_{0}$. The theoretical
curves are calculated from the matter power spectrum corresponding
to the parameters of the simulation using Equation~(\ref{eq:TheoryVmu}).
As $c_{2}^{0}$ corresponds to $\sigma^{2}$, we also plot the result
of a standard estimation of $\sigma^{2}$ for the observed galaxies.
\protect \\
Bottom four: the same quantities but all divided by the theoretical
prediction of the LasDamas power spectrum. The value of the straight line indicates the bias between
the linear correlation function and the correlations in the ``bright
sample''. The line following the trend of the SDSS data corresponds
to the theoretical prediction for a power spectrum with the {\em Planck}
parameters $\Omega_{\rm m}\approx0.32$. The $\chi^2$ values are given in Table~\ref{tab:brightcsigma}.%
\label{fig:CorrInt-c2}}
\end{figure*}

As the resampling of the points introduces an important amount of
correlation between the realizations and as the $r$--dependence of
the $v_{\mu}$ also contains some correlation, we use the $160$ mock
samples to estimate the covariance matrix for the points of the empirical
function $\tilde{\overline{V}}_{\mu}\left(r,\varrho_{0}\right)$.
We use $\tilde{\overline{V}}_{\mu}\left(\varrho_{0}\right)$ at four
different distances $r$ and use its values at $20$ of the $25$
$\varrho_{0}$ points. This gives us a covariance matrix of dimension
$80\times80$. This is still quite large, given that we only have $160$ mock samples: there is an important uncertainty in the estimation of the covariance matrix. In addition to the use of the unbiased estimator of \cite{2007A&A...464..399H}, as in Section~\ref{sub:The-functionals}, we, therefore, need to take into account the propagation of the errors in the estimated covariance matrix to the values we want to measure. For doing so, we use the prescription in \cite{2014MNRAS.439.2531P}. This considerably increases our errorbars.
We use the resulting covariance matrix for a polynomial
fit of fourth order to the average $\tilde{\overline{V}}_{\mu}\left(r,\varrho_{0}\right)$
for all the mock samples and for the SDSS. With Equation~(\ref{eq:MinkIntegr})
this gives us the coefficients $b_{2}$--$b_{4}$ for both the mock samples
and the SDSS. 

We finally divide these coefficients by the corresponding power of the
volume and arrive at an estimate of the integrals:
\begin{eqnarray}
c_{n+1}^{\mu} & = & \frac{b_{n+1}^{\mu}}{V_{0}^{n+1}\left(B\right)}=\frac{1}{V_{0}^{n+1}\left(B\right)}\int_{\cD}\d^{3}x_{1}\dots\d^{3}x_{n}\label{eq:NormInteg}\\
 &  & \times\xi_{n+1}\left(0,\bx_{1},\dots\bx_{n}\right)V_{\mu}\left(B\cap B_{\bx_{1}}\cap\dots\cap B_{\bx_{n}}\right),\nonumber 
\end{eqnarray}
where $V_{0}\left(B\right)$ is the volume of a Ball of radius $R$.

\subsection{Integrals of the two--point correlation function\label{sub:Integrals-two}}

The results for the coefficients $c_{2}^{\mu}$, i.e. those involving
an integral over the two--point correlation function, are shown in
the upper four plots of Fig.~\ref{fig:CorrInt-c2}. As described
in Section~\ref{sub:Gauss-Poisson-process}, the first quantity plotted,
$c_{2}^{0}$, is related to the matter variance in spheres of radius
$R$. In fact, from Equation~(\ref{eq:NormInteg}) and~(\ref{eq:V0-sigma}),
one can see that $c_{2}^{0}=\sigma^{2}$. So, the first plot of Figure
\ref{fig:CorrInt-c2} compares different ways of calculating $\sigma^{2}$.
The data and mock points and their error bars are derived from the
MFs using the method just described. The points
of $\sigma_{\rm emp.}^{2}$ are calculated from the observed sample with
the usual estimator for the matter variance in spheres of radius $R$,
\begin{equation}
\sigma^{2}\left(R\right)=\frac{\BE\left[M\left(R\right)^{2}\right]-\BE\left[M\left(R\right)\right]^{2}}{\BE\left[M\left(R\right)\right]^{2}}\;,\label{eq:SigDef}
\end{equation}
where $M\left(R\right)$ is the integrated matter density of the sphere
and $\BE\left[X\right]$ is the average over all spheres.

The third way of calculating $\sigma^{2}$ is direct integration of
the theoretical power spectrum using Equation~(\ref{eq:TheoryVmu}) which,
for $c_{2}^{0}$, directly becomes Equation~(\ref{eq:sigmaPower}).
For the form of $P\left(k\right)$, we use the parametrization of
\cite{1998ApJ...496..605E} that includes the effects
of baryons. However, we use the form without their oscillations. We
normalize the amplitude of the power spectrum to match the
amplitude of the mocks. This gives a linear bias of $b\approx2.37$,
i.e. $\sigma_{\rm mock}^{2}=b^{2}\sigma_{\rm lin.}^{2}$.

In the $V_{1}$--$V_{3}$ plots of Fig.~\ref{fig:CorrInt-c2}, we
use the power spectrum with the same normalization and Equation~(\ref{eq:TheoryVmu})
for the theory prediction.
The first plot of Fig.~\ref{fig:CorrInt-c2} shows that all three
ways of calculating $\sigma^{2}$ are overall in agreement.
This is also true for the plots derived from $\ol{V}_{1}$--$\ol{V}_{3}$,
but there is also a deviation of the coefficients $c_{2}^{\mu}$
of the observed galaxies from those of the simulated mock galaxies.
On small scales they start quite close, but on larger scales the deviation
becomes more important. It is interesting to see that the mocks are indeed
well--described by the theoretical power spectrum that entered into their
simulation. This shows that the simulations and the extraction procedure
give a consistent picture. The observed galaxies, however,
deviate slightly from the simulated cosmology. This means that, even
though the overall normalization of the correlation function is correct
by the consistency of the $\sigma^{2}$ results, other features of
the correlation function are not captured equally well.

The lower four plots of Fig.~\ref{fig:CorrInt-c2} show the same
quantities, but divided by the theoretical prediction for the
simulated cosmology. This removes the general trend and allows
a more detailed comparison. Thus, in these plots, the scales on the
y--axis have only a relative meaning: the value $y=1$ marks the (scale--dependent) 
integrated mock power spectrum for $\Omega_{\rm m}=0.25$.
The value of $\approx5.6$ for the straight mock line corresponds to the squared bias of the mocks $b^2$.
The line going through the points of the SDSS data, represents the $c^{\mu}_{2}$ calculated from a {\em Planck} 
$\Omega_{\rm m}\approx0.32$ power spectrum for comparison.
Fig.~\ref{fig:CorrInt-c2} shows that the mild deviations
of the MFs for the data and the mock galaxies, as
found in Figs~\ref{fig:bright-av} and \ref{fig:bright-min},
already occur at the level of the first correction to the leading
Poisson term in the expansion $\ol V_{\mu}$, Equation~(\ref{eq:MinkCorrCon}) (but they are a bit less significant here without the higher order information, see Table~\ref{tab:brightcsigma}).
We shall turn to the influence of the higher orders in the
next section.
\begin{table}
\center
\begin{tabular}{p{0.02\linewidth}p{0.04\linewidth}p{0.04\linewidth}|p{0.04\linewidth}p{0.04\linewidth}p{0.02\linewidth}p{0.04\linewidth}p{0.04\linewidth}p{0.04\linewidth}|p{0.04\linewidth}p{0.05\linewidth}}
\cline{1-5} \cline{7-11} 		\noalign{\smallskip} 		

 & $\chi^2$ & $\sigma_G$ & $\chi^2_I$ & $\sigma_{G\,I}$& & & $\chi^2$ & $\sigma_G$ & $\chi^2_I$ & $\sigma_{G\,I}$\\
\noalign{\smallskip} 		\cline{1-5} \cline{7-11} 		\noalign{\smallskip}

 $c^0_{2}$ & 11.4 & 2.30 & 16 & 3.0 & & $c^0_{2}$ & 8.24 & 1.73 & 3.7 & 0.76 \\
 $c^1_{2}$ & 5.11 & 1.09 & 12 & 2.4 & & $c^1_{2}$ & 0.75 & 0.07 & 0.25 & 0.01 \\
 $c^2_{2}$ & 8.84 & 1.84 & 10 & 2.0 & & $c^2_{2}$ & 1.34 & 0.18 & 0.15 & 0.003 \\
 $c^3_{2}$ & 3.22 & 0.64 & 7.7 & 1.6 & & $c^3_{2}$ & 0.35 & 0.02 & 0.14 & 0.003 \\

\noalign{\smallskip} 	\cline{1-5} \cline{7-11} 	\noalign{\smallskip}

\end{tabular}%
\caption{$\chi^2$ values for the SDSS data points in the lower panel of Fig.~\ref{fig:CorrInt-c2} from the theoretical curves. The table on the left compares the (red) SDSS points to the straight line of the LasDamas cosmology. The table on the right compares them to the dashed line of the {\em Planck} cosmology. The $\chi^2$ values are over $4$ degrees of freedom.}
\label{tab:brightcsigma} 
\end{table}

\subsection{Integrals of the three--point correlation function\label{sub:Integrals-three}}

In addition to the integrals over the two--point correlation function
of the previous section, which are completely describing a Gauss--Poisson
point process, also the higher order terms are important for a general point distribution.
In this sense, the first corrections to the Gaussian point distribution
are the integrals $c_{3}^{\mu}$.

\begin{figure*}
\includegraphics[width=0.9\textwidth]{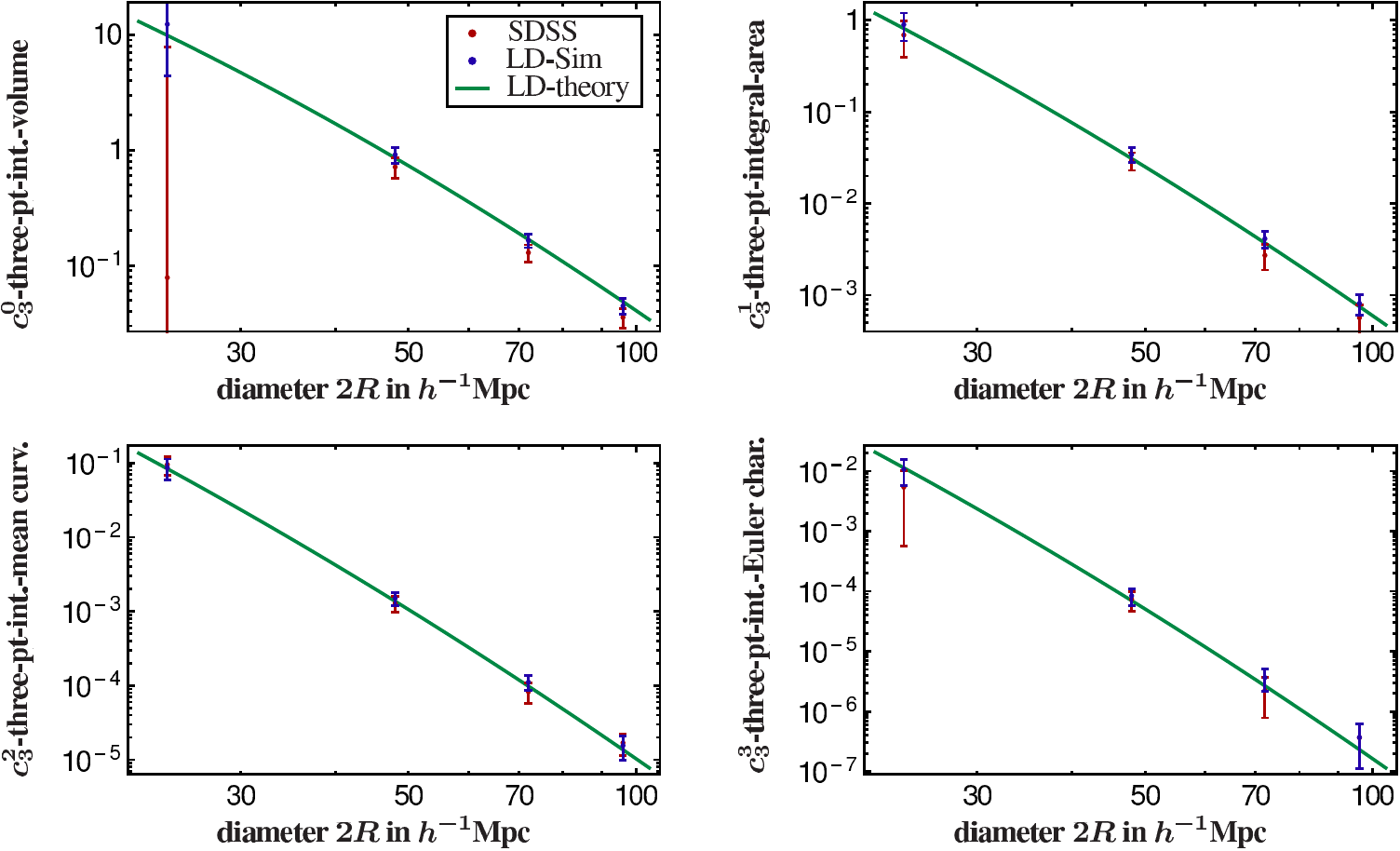}
\includegraphics[width=0.9\textwidth]{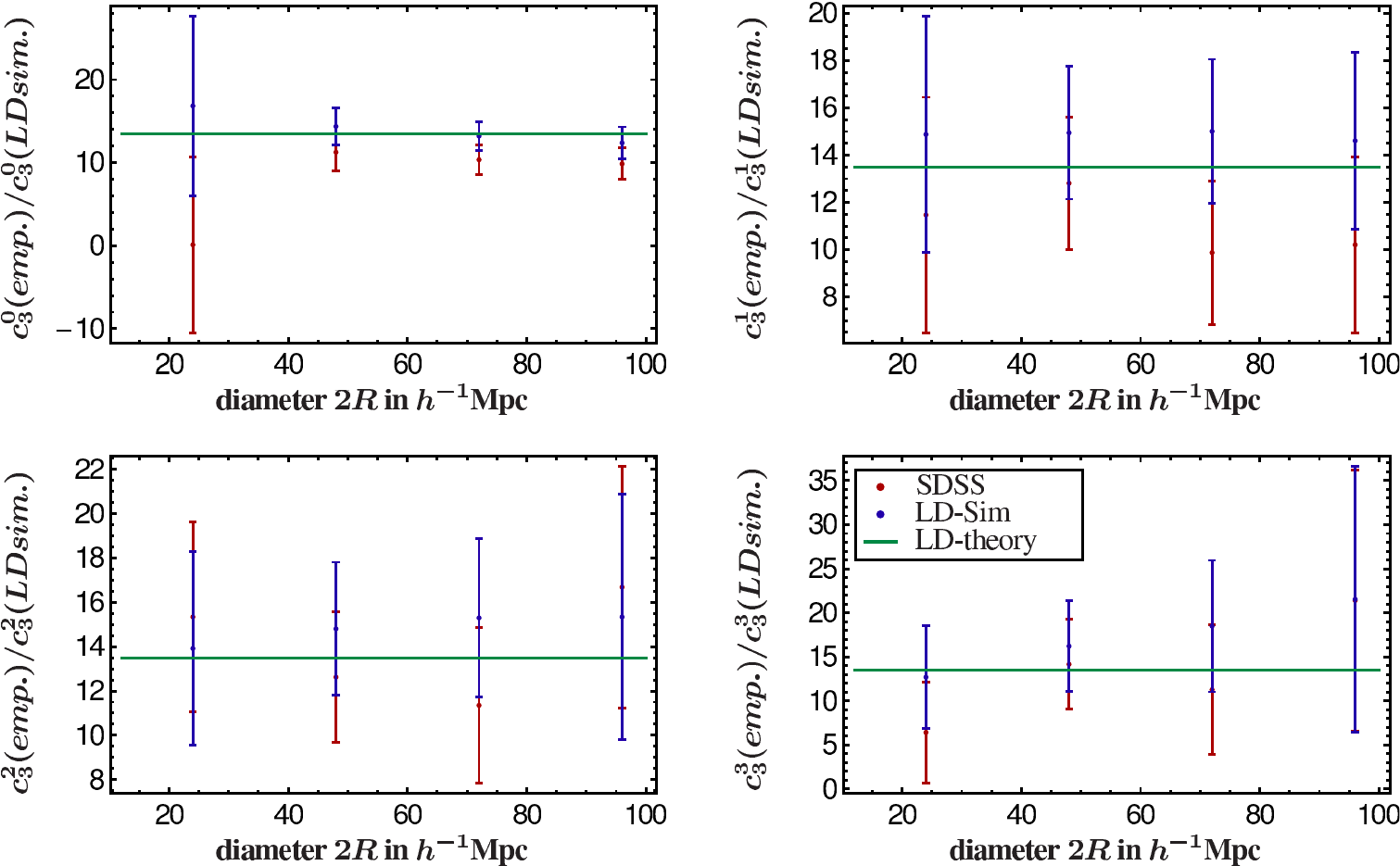}
\caption{
Top four: The four Minkowski integrals of the three--point correlation function,
i.e. $c_{3}^{\,\mu}$ from Equation~(\ref{eq:NormInteg}), for the SDSS
LRG `bright sample' (lower points, in red) and the average of the corresponding mock samples (upper points, blue).
The error bars are the $1\sigma$ errors from the diagonal of the covariance matrix of
the fit of $\tilde{\overline{V}}_{\mu}\left(\varrho_{0}\right)$ with
a polynomial of fourth order in $\varrho_{0}$.
The lines are the integrals of Equation~(\ref{eq:ThreePtInteg}) evaluated for the LasDamas cosmology. \protect \\
Bottom four: the same quantities but all divided by the theoretical
prediction of the LasDamas power spectrum. The value of the straight line indicates the (linear) bias between
the linear three--point correlation function and the correlations in the `bright sample'.}
\label{fig:CorrInt-c3}
\end{figure*}

\begin{figure*}
\includegraphics[width=0.93\textwidth]{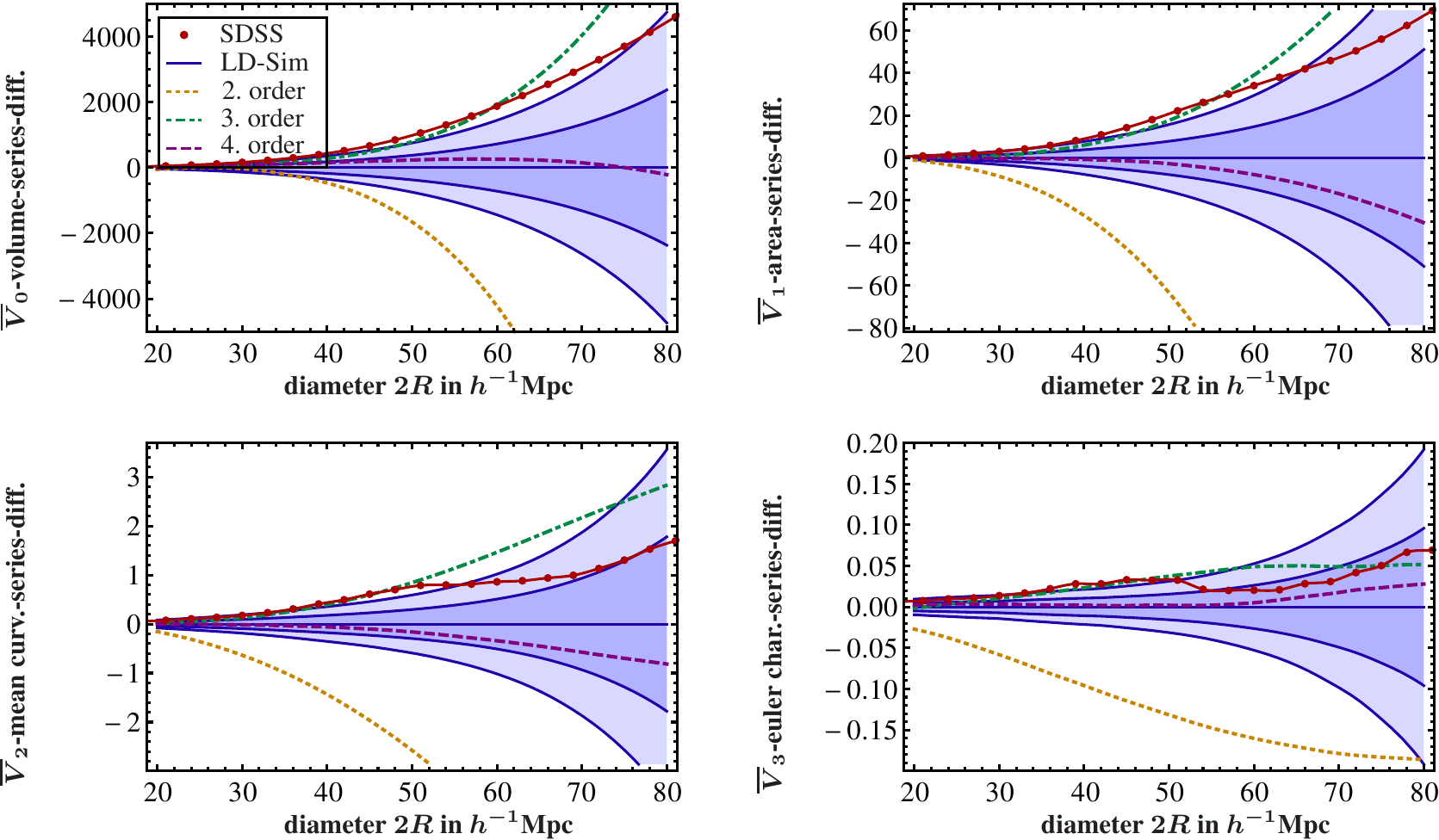}
\caption{The four modified MFs $\ol V_{\mu}$ of 30\% of
the `bright sample', after subtraction of the average of the mocks.
The (yellow) dotted line shows the theoretical expectation for a pure Gauss--Poisson
process of this density and this amount of two--point correlation.
The (purple) dashed line gives the functionals as derived from the expansion
(\ref{eq:MinkCorrCon}) truncated at $\varrho_{0}^{3}$ with the coefficients
$c_{2}^{\mu}$--$c_{4}^{\mu}$ as found in Sections~\ref{sub:Integrals-two}
and~\ref{sub:Integrals-three}.\label{fig:blow-min}}
\end{figure*}

These integrals $c_{3}^{\mu}$ are shown in Fig.~\ref{fig:CorrInt-c3}.
In comparison with Figure
\ref{fig:CorrInt-c2} we recognize that the amplitude of the $c_{3}^{\mu}$
integrals is larger by a factor of $5$ than the integrals $c_{2}^{\mu}$.
Like in Fig.~\ref{fig:CorrInt-c2}, we also have a slight but again less significant 
deviation for these quantities.

The lines in these plots are calculated from a tree--level bispectrum using Equation~(\ref{eq:ThreePtInteg}) and the bias of $b=2.37$ as found from the two point normalization.

As described in the beginning of this section, we use a fit up to
$\varrho_{0}^{4}$ and, therefore, we have also determined the coefficients
$c_{4}^{\mu}$. However, the quality of the determination of those
coefficients becomes even worse than already for the $c_{3}^{\mu}$
coefficients and therefore we do not plot them here.

\subsection{Importance of higher order correlations\label{sub:Importance-of-higher}}

To get an impression of how well the first four terms
in the series~(\ref{eq:MinkCorrCon}) already describe the MFs
of the `bright' galaxy sample, we use the coefficients
$c_{2}^{\mu}\left(R\right)$--$c_{4}^{\mu}\left(R\right)$, as obtained
from the fit to $\tilde{\overline{V}}_{\mu}\left(\varrho_{0}\right)$
in the two previous sections, to calculate the $\ol V_{\mu}$ of
Equation~(\ref{eq:MinkCorrCon}) up to $n=3$.

This decomposition allows us to address another interesting question: how non--Gaussian the point distribution actually is. For a pure Gauss--Poisson point process, we have seen in Section \ref{sub:Gauss-Poisson-process} that a truncation after $n=1$ is exact. The comparison of the contribution of $c_{2}^{\mu}\left(R\right)$ with the other components $c_{n+1}^{\mu}\left(R\right)$ provides a measure of non-Gaussianity.
This is only a meaningful test, however, if the distribution can be Gaussian in the first place. Due to the positive definite matter density field $\varrho$ there are some restrictions, as discussed around Equation~(\ref{eq:xicond}). For a Gauss--Poisson process
to exist, the density must be low enough for a given amount of two--point
correlation. The precise relation between $\varrho_{0}$ and $\xi_{2}$ is that of Equation~(\ref{eq:xicond}):
\[
\varrho_{0}\int_{\cD}d\by\xi_{2}\left(\left|\by\right|\right)\le1\;.
\]
For the strongly clustered SDSS LRGs this requirement is not fulfilled for the full density of the sample.
Also for 80\% of the density like in Figs~\ref{fig:bright-av} and
\ref{fig:bright-min}, where the density was $\varrho_{0}=2.1\times10^{-5}\hDens$, the two point amplitude is 
too high to allow for a Gaussian approximation. So already from this condition, we know that the point sample is not a Gaussian distribution.
However, for a sample having 30\% of the full density, i.e. $\varrho_{0}=0.78\times10^{-5}\hDens$,
the condition is (marginally) satisfied. Therefore, we calculated
the modified MFs for a large number of realizations of 30\%
of the `bright sample' mocks in the same way as in Section~\ref{sub:The-functionals}.

Fig.~\ref{fig:blow-min} shows the influence of the first three components $c_{n+1}^{\mu}\left(R\right)$ on the series of the modified MFs $\ol V_{\mu}$. We confirm that the mock point catalogues are not a realization of a Gauss--Poisson process and that higher order corrections are crucial for the MFs.

With the inclusion of $c_{3}^{\mu}$ and $c_{4}^{\mu}$, this truncated
series is quite good in describing the modified MF
densities of the data up to a scale of around $60\hMpc$. For larger
scales, the coefficients $b_{n+1}^{\mu}$ become too big (even though
the $c_{n+1}^{\mu}$ decay with $R$, they decay slower than $V_{0}^{n+1}\left(B\right)$
and, therefore, the $b_{n+1}^{\mu}$ grow), and the quality of the
approximation becomes worse. This deviation from the approximated
function shows that, even for densities
as low as the present $\varrho_{0}=0.78\times10^{-5}\hDens$, the MFs include
contributions from galaxy correlations way beyond the standard two--point
correlations. This confirms the claim made in the introduction that
they are sensitive to higher order correlations. For higher densities these contributions become even more important as $\ol V_{\mu}$, Equation~(\ref{eq:MinkCorrCon}), is a power series in $\varrho$.

\subsection{Coefficients of a lognormal distribution}
\label{sec:lognorm}
The non--Gaussianity investigated in the previous section is, of course, not surprising. It is well known that the non--linear evolution of the density field leads to deviations from the initial Gaussian distribution. A better model is provided by a lognormal distribution, proposed by \citet{1991MNRAS.248....1C} and investigated in \citec{1994ApJ...435..536C,2011ApJ...735...32W}. Comparing it to simulations, it has been shown to reproduce well the one-point function \citep[see e.g. ][]{1994ApJ...420...44K} and has been used recently to enhance the extraction of information from the two-point function \citep{2009ApJ...698L..90N,2012ApJ...748...57S}. In comparison with a Gaussian distribution, however, it has the inconvenience not to be described by a simple truncation of the $\ol V_{\mu}$ series. This is why we concentrated on the Gaussian reference case in the previous section. Nevertheless, the lognormal distribution has the interesting property that, even though it contains an infinite number of higher order correlations, they are not independent of the two--point properties; the knowledge of the two--point function fixes the form of the higher order terms. We shall exploit this property in this section to compare the $c_{n+1}^{\mu}$ coefficients we extracted from the mock samples with the lognormal prediction.

\begin{figure}
\includegraphics[width=0.45\textwidth]{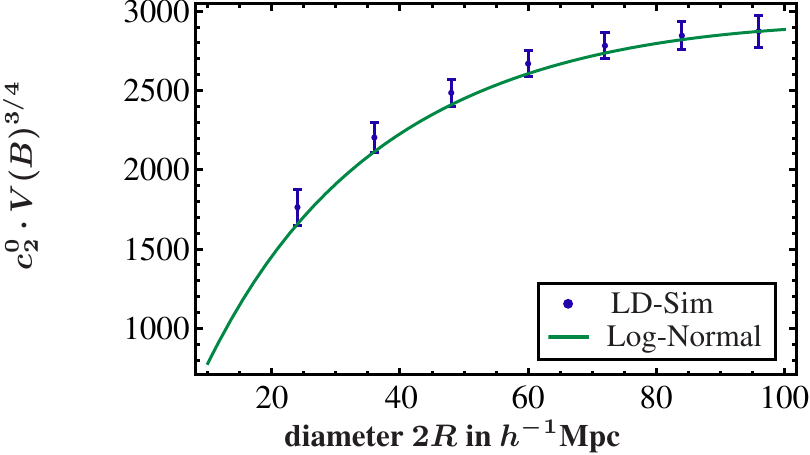}\hspace{4em}\includegraphics[width=0.45\textwidth]{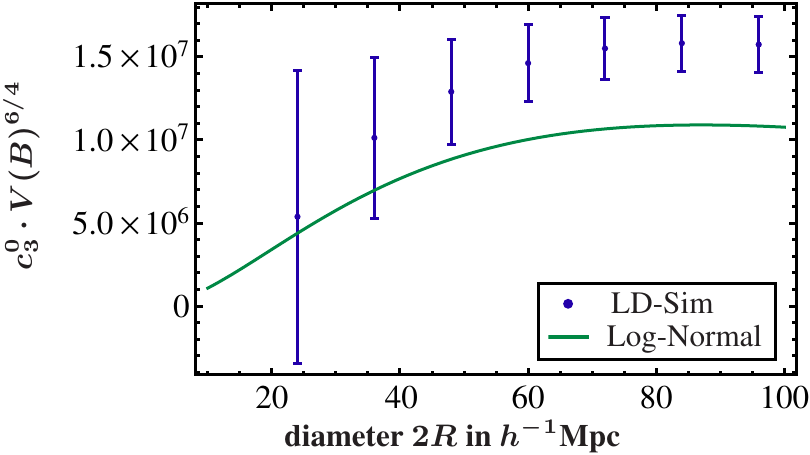}
\includegraphics[width=0.45\textwidth]{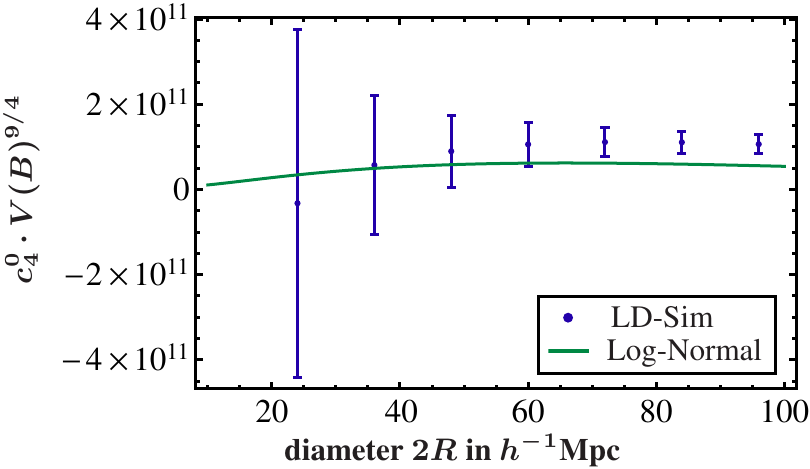}
\caption{Comparison of the coefficients $c^{0}_{2}$--$c^{0}_{4}$ defined in equation~(\ref{eq:NormInteg})
to those of a lognormal distribution with the same two--point
function. For a better visibility, the coefficients $c^{0}_{n}$ have been multiplied by $V\left(B\right)^{3/4(n-1)}$ in these plots. 
\label{fig:LogNormal}}
\end{figure}

We first have to choose the two--point properties for which we use the standard two--point correlation function for the LasDamas cosmology. Then, we can calculate, in principle, all higher connected correlation functions for a multivariate lognormal distribution.
In practice, however, this quickly leads to long expressions, a combinatorial increase in the number of two--point functions. Therefore, we only use connected correlation functions up to $\xi_{4}$.

Using the $\xi$s constructed in this way, we can evaluate the series
(\ref{eq:MinkCorrCon}) up to an index $n=4$. However, we restrict ourselves to $V_{0}$, because the necessary integral can be expressed for any $n$ in a rather compact way.

Fig.~\ref{fig:LogNormal} then shows the result. The two--point contribution matches quite well, which is not surprising as it is the same as in the previous case. The integrals over the higher order correlation functions, however, deviate quite strongly from the data measured from the mock samples. This shows that higher order correlations can also be crucial to test improved reference models.

\section{Conclusion}
\label{sec:Conclusion}

In this paper, we provided new analysis tools based on the Germ--Grain model to calculate the 
family of Minkowski Functionals of point sets. The code is made available to the community with this
paper that complements the existing Germ--Grain codes. The advantage of the former codes is 
still the possibility to deconvolve the boundaries, which is especially needed for sparse catalogues with
complicated mask structure. The advantage of the new code has to be seen in the fast performance
for large samples, the possibility to study the Partial Minkowski Functionals locally
and in explicit relation to correlation properties of the data sets.

As a first application of the new code,
we also performed an extensive analysis of the Minkowski
Functionals of the SDSS LRG sample for the Germ--Grain model. We favour
this model because it provides a direct way to analyse the data without
additional smoothing, and because it provides analytical relations
between the Minkowski Functionals and the $N$--point correlation
functions of the galaxy distribution. We especially had a detailed look at those correlation properties
to locate the deviations found between the mock samples and the SDSS data.

In Section~\ref{sec:Minkowski-fun-LRG}, we compared the Minkowski
Functionals of the observed galaxies to a grid of $\Lambda$CDM $N$--body
simulations of the galaxy distribution, for two different luminosity
thresholds. In both cases, the Minkowski Functionals of the observed
and the simulated galaxy distributions show significant disagreement.
For the galaxies with the higher luminosity, this difference is statistically less
important than for the `dim' LRG sample.

Making use of the analytic relation to the correlation functions provided in this paper,
we derived in Section~\ref{sec:Non-Gaussian-correlations} some integrals
over the two-- and three--point function of the galaxy distribution.
We compared the results to
the prediction for structure in a perturbed Friedmannian universe. We
found that this prediction describes quite well the simulated galaxy
distribution, but fails to describe the observed one.

We also showed that the galaxy distribution is clearly different from
a Gauss--Poisson distribution (and for higher orders also from a lognormal distribution), which demonstrates how higher order correlations
are crucial to describe the observed structure. As previously often emphasized, it is
necessary to address higher order correlations for the purpose of determining
morphological fluctuations. It is not sufficient, if the density or
the two--point measures agree on selected samples, in order to conclude on
the absence of significant fluctuations or the reality of large structures.

It will be interesting to see, which aspects of these results will become more significant in larger and deeper surveys.

\section*{Acknowledgements}

We thank Martin Kerscher, Carlo Schimd, Francesco Sylos Labini and Herbert Wagner for helpful discussions.
AW thanks Lars Andersson, Steffen Hess and Cameron McBride for interesting remarks.
The work of AW was partially supported by the DFG under grant GRK
881. The work by TB was conducted within the `Lyon Institute of
Origins', Grant No. ANR--10--LABX--66. The Minkowski code \textsc{\footnotesize CHIPMINK} is a  
completely revised version by MO, with additions by AW, based on a former code architecture
(including a parallel architecture) written by Jens Schmalzing and Alexander Rabus. 
TB and MO acknowledge many fruitful discussions with them.
The \textsc{\footnotesize MINKOWSKI--4} package is available on request by sending email to TB.
\vspace{0.1in}

The simulations used for the Minkowski Functional analysis were carried out by the Large Suite of Dark Matter Simulations (LasDamas) project. The data are publicly available at \href{http://lss.phy.vanderbilt.edu/lasdamas/}{http://lss.phy.vanderbilt.edu/lasdamas/}.
\vspace{0.1in}

Funding for the Sloan Digital Sky Survey (SDSS) has been provided by the Alfred P. Sloan Foundation, the Participating Institutions, the National Aeronautics and Space Administration, the National Science Foundation, the US Department of Energy, the Japanese Monbukagakusho, and the Max Planck Society. The SDSS website is \href{http://www.sdss.org/}{http://www.sdss.org/}.

The SDSS is managed by the Astrophysical Research Consortium (ARC) for the Participating Institutions. The Participating Institutions are The University of Chicago, Fermilab, the Institute for Advanced Study, the Japan Participation Group, The Johns Hopkins University, Los Alamos National Laboratory, the Max-Planck-Institute for Astronomy (MPIA), the Max-Planck-Institute for Astrophysics (MPA), New Mexico State University, University of Pittsburgh, Princeton University, the United States Naval Observatory, and the University of Washington.

\bibliographystyle{mn2e_eprint}

\appendix

\section{Derivation of Equation~(3)}

\label{sec:derivation}

As we make heavy use of the analytic relation~(\ref{eq:MinkDensDef}),
we recall here the derivation of this formula in the general formulation
that we need here. The version for the Poisson case has been shown
in \cite{mecke1991euler}. \cite{phdSchmalzing} and \cite{1999MNRAS.309.1007S} state the
general case, but only sketch the derivation.

We shall prove equation~(\ref{eq:MinkDensDef}) for a statistically homogeneous
point process on a three--dimensional support $\cD\in\mathbb{E}^{3}$.
The key ingredients for the proof will be the additivity and motion
invariance of the MFs, the principal kinematical
formula and statistical homogeneity (i.e. the motion invariance of
statistical quantities characterizing the point process).

To derive the relation for all four MFs in one go,
we first define the Minkowski polynomial as follows: 
\begin{equation}
M\left(t;K\right)=\sum_{\mu=0}^{3}\frac{t^{\mu}}{\mu!}\alpha_{\mu}V_{\mu}\left(K\right),
\end{equation}
with $\alpha_{\mu}=\frac{\omega_{d-\mu}}{\omega_{d}}$ and $\omega_{j}=\frac{\pi^{j/2}}{\Gamma\left(j/2+1\right)}$.
To recover the functionals we can take the derivatives 
\begin{equation}
V_{\mu}\left(K\right)=\frac{1}{\alpha_{\mu}}\left.\frac{\partial^{\mu}M\left(t;K\right)}{\partial t^{\mu}}\right|_{t=0}\;.
\end{equation}
These Minkowski polynomials obey the additivity relation of the Minkowski
Functionals, $\text{ \ensuremath{\forall}\ }\CB_{1},\CB_{2}\in\CR$
($\CR$ is the ring of polyconvex bodies): 
\begin{equation}
M\left(\CB_{1}\cup\CB_{2}\right)=M\left(\CB_{1}\right)+M\left(\CB_{2}\right)-M\left(\CB_{1}\cap\CB_{2}\right)\;.
\end{equation}
So, for a collection of $N$ Balls located at the positions $r_{i}$
[in short $B\left(\br_{i}\right)=B_{i}$ in the following], we have
\[
M\left(\cup_{i=1}^{N}B_{i}\right)=\sum_{i=1}^{N}M\left(B_{i}\right)-\sum_{i<j}M\left(B_{i}\cap B_{j}\right)+\dots
\]
\begin{equation}
+\left(-1\right)^{N+1}M\left(B_{1}\cap\dots\cap B_{N}\right).
\end{equation}
where $M\left(B_{1}\cap\dots\cap B_{N}\right)$ is the Minkowski polynomial
of the intersection of the $N$ Balls, which is $0$ if they do not
intersect. 

Equation~(\ref{eq:MinkDensDef}) describes the statistical average
over an ensemble of such realizations of the point process. To calculate
these average MFs for a point process described
by a given set of $n$--point correlation functions up to $n=N$,
we weight such a configuration of Balls with its probability 
\begin{equation}
p\propto\varrho_{N}\left(x_{1},x_{2},\dots,x_{N}\right)dV_{1}dV_{2}\dots dV_{N}\;.
\end{equation}
$\varrho_{N}\left(x_{1},x_{2},\dots,x_{N}\right)=\av{\varrho\left(x_{1}\right)\dots\varrho\left(x_{N}\right)}$
is the complete $N$--point correlation function that is related to
the probability of finding particles at the $N$ positions $x_{n}$
simultaneously. Abbreviating the integration measure by 
\begin{equation}
\int_{\cD}\d\tau_{n}=\int_{\cD}\d^{3}x_{1}\d^{3}x_{2}\dots \d^{3}x_{n}\;,
\end{equation}
we, therefore, find for the average 
\[
\av M=\frac{1}{N^{N}}\int_{\cD}\d\tau_{N}\varrho_{N}\left(x_{1},x_{2},\dots,x_{N}\right)M\left(\cup_{i=1}^{N}B_{i}\right)
\]
\[
=\sum_{n=1}^{N}\left(\begin{array}{c}
N\\
n
\end{array}\right)N^{-n}\left(-1\right)^{n+1}\int_{\cD}\d\tau_{n}\varrho_{n}\left(x_{1},x_{2},\dots,x_{n}\right)
\]
\begin{equation}
\times M\left(B_{x_{1}}\cap B_{x_{2}}\cap\dots\cap B_{x_{n}}\right).
\end{equation}
In the limit of an infinite structure $N\rightarrow\infty$, $\left(\begin{array}{c}
N\\
n
\end{array}\right)N^{-n}\rightarrow\frac{1}{n!}$, and so 
\begin{multline}
\av M=1-\sum_{n=0}^{\infty}\frac{\left(-1\right)^{n}}{n!}\int_{\cD}\d\tau_{n}\varrho_{n}\left(x_{1},x_{2},\dots,x_{n}\right)\times\\
M\left(B_{x_{1}}\cap B_{x_{2}}\cap\dots\cap B_{x_{n}}\right),\label{eq:avMinkstat}
\end{multline}
which is a result already obtained in \cite{1994A&A...288..697M} for the individual functionals. 

To pass from this expression which involves the complete $N$--point
correlations to the formulation in~(\ref{eq:MinkDensDef}) with the
connected $N$--point correlation functions $\xi_{n}$ we need three
ingredients:
\begin{enumerate}
\item the principal kinematical formula%
\begin{equation}
\int_{\cD}\d^{3}xM\left(A\cap B_{x}\right)\sim_{{\rm mod}\left(t^{4}\right)}M\left(A\right)M\left(B\right),\label{eq:principalkinematical}
\end{equation}%
that connects the integral of the Minkowski polynomial of the intersection
$A\cap B_{x}$ with a product of Minkowski polynomials when dropping
all terms involving powers of $t$ larger than $t^{3}$. Here $A,B\in\CR$ and it is important that there is no function of $x$ in the integral.
\item the motion invariance of the MFs
\[
M\left(A_{\bx}\cap B_{\bx+\by}\right)=M\left(A_{\mathbf{0}}\cap B_{\by}\right)
\]

\item the motion invariance of the correlation functions of a statistically
homogeneous distribution
\[
\xi_{n}\left(\by,\by+\bx_{1},\dots,\by+\bx_{n-1}\right)=\xi_{n}\left(\mathbf{0},\bx_{1},\dots\bx_{n-1}\right)
\]

\end{enumerate}
With these ingredients we can now show the following formal cumulant
relation
\begin{equation}
1+\sum_{n=1}^{\infty}m_{n}t^{n}/n!=\exp\left(\sum_{n=1}^{\infty}\kappa_{n}t^{n}/n!\right),\label{eq:formalcumulant}
\end{equation}
connecting the `moments'
\begin{equation}
m_{n}=\frac{1}{V_{\cD}}\int_{\cD}\d\tau_{n}\varrho_{n}\left(x_{1},\dots,x_{n}\right)M\left(B_{x_{1}}\cap\dots\cap B_{x_{n}}\right)\label{eq:defmn}
\end{equation}
to the `cumulants'
\begin{equation}
\kappa_{n}=\frac{1}{V_{\cD}}\int_{\cD}\d\tau_{n}\varrho^{n}\xi_{n}\left(x_{1},\dots,x_{n}\right)M\left(B_{x_{1}}\cap\dots\cap B_{x_{n}}\right)\;.\label{eq:defxin}
\end{equation}
The relation (\ref{eq:formalcumulant}) is satisfied iff
\[
m_{n}=\sum_{\pi}\prod_{B\in\pi}\kappa\left(X_{i}:i\in B\right)
\]
where $\pi$ is the set of all partitions of $\left\{ 1,\dots,n\right\} $.
We already know that there is such a relation connecting $\varrho_{n}$
and the $\xi_{n}$, because the latter are the joint cumulants of
the former, up to a factor involving the background density $\varrho=\varrho_{1}\left(\bx_{1}\right)=\av{\varrho\left(\bx_{1}\right)}$
which is nonzero and constant on the support $\cD$:
\[
\varrho_{n}\left(x_{1},x_{2},\dots,x_{n}\right)=\sum_{\pi}\prod_{B\in\pi}\varrho^{n}\xi\left(\varrho\left(x_{i}\right):i\in B\right)
\]
To show that $m_{n}$ and $\kappa_{n}$ as defined above also fulfil
this condition, we do a weighted integral on both sides
\begin{eqnarray}
\frac{1}{V_{\cD}}\int_{\cD}\d\tau_{n}\varrho_{n}\left(x_{1},\dots,x_{n}\right)M\left(B_{x_{1}}\cap\dots\cap B_{x_{n}}\right)=\;\;\;\;\;\;\;\;\;\;\;\;\;\;\; & & \label{eq:cumulantintegral}\\
\sum_{\pi}\prod_{B\in\pi}\frac{1}{V_{\cD}}\int_{\cD}\d\tau_{n}\varrho^{n}\xi\left(\varrho\left(x_{i}\right):i\in B\right)M\left(B_{x_{1}}\cap\dots\cap B_{x_{n}}\right) & & \nonumber
\end{eqnarray}
and find $m_{n}$ on the left--hand side. What we need to demonstrate is
that the products of $\xi_{i}$s become products of the related $\kappa_{i}$s.
This means that we have to break up the term $M\left(D\cap B_{x_{1}}\cap B_{x_{2}}\cap\dots\cap B_{x_{n}}\right)$
into products of $M$s corresponding to the $\xi$s. We show how
this is done for one generic product.%

Let $B\in\pi$ be the partition $\left\{ \left\{ 1,\dots,i\right\} ,\left\{ i+1,\dots,n\right\} \right\} $
and $j=n-i$, then one of the products in equation (\ref{eq:cumulantintegral})
consists of two terms which are
\begin{eqnarray}
\frac{1}{V_{\cD}}\int_{\cD}\d\tau_{n}\xi_{i}\left(x_{1},\dots,x_{i}\right)\xi_{j}\left(x_{i+1},\dots,x_{n}\right)\times\;\;\;\;\;\;\;\;\;\;\;\;\;\;\; & &\nonumber\\
M\left(B_{x_{1}}\cap B_{x_{2}}\dots\cap B_{x_{i}}\cap B_{x_{i+1}}\cap B_{x_{i+2}}\cap\dots\cap B_{x_{n}}\right)\;.\nonumber
\end{eqnarray}
We can now make use of our assumption of a statistically homogeneous
point distribution and make a change of coordinates from $x_{1},\dots,x_{i}$
to $x_{1},y_{1}\dots,y_{i-1}$ where the new coordinates $y_{k}$
are simply $y_{k}=x_{k+1}-x_{1}$ 
\begin{align*}
&\frac{1}{V_{\cD}}\int_{\cD}\d^{3}x_{1}\d^{3}y_{1}\dots \d^{3}y_{j-1}\xi_{i}\left(y_{1},\dots,y_{i-1}\right)\int_{\cD}\d\tau_{j}\xi_{j}\left(x_{i+1},\dots,x_{n}\right)\times\\
&M\left(B_{x_{1}}\cap B_{x_{1}+y_{1}}\dots\cap B_{x_{1}+y_{i-1}}\cap B_{x_{i+1}}\cap B_{x_{i+2}}\cap\dots\cap B_{x_{n}}\right)
\end{align*}
Due to statistical homogeneity $\xi_{i}$ does not depend on $x_{1}$
and we can use the principal kinematical formula (\ref{eq:principalkinematical})
to split the $M$ into 
\begin{align*}
&\frac{1}{V_{\cD}}\int_{\cD}\d^{3}y_{1}\dots \d^{3}y_{j-1}\xi_{i}\left(y_{1},\dots,y_{i-1}\right)M\left(B_{0}\cap B_{y_{1}}\dots\cap B_{y_{i-1}}\right)\\
&\times\int_{\cD}\d\tau_{j}\xi_{j}\left(x_{i+1},\dots,x_{n}\right)M\left(B_{x_{i+1}}\cap B_{x_{i+2}}\cap\dots\cap B_{x_{n}}\right)\;.
\end{align*}%
Formally, we can reinsert the integral over $x_{1}$, $\frac{1}{V_{\cD}}\intop \d^{3}x_{1}$,
and when we change the variables back to the redundant $x_{1},\dots,x_{i}$
we have
\begin{align*}
&\frac{1}{V_{\cD}}\int_{\cD}\d\tau_{i}\xi_{i}\left(x_{1},\dots,x_{i}\right)M\left(B_{x_{1}}\cap B_{x_{2}}\dots\cap B_{x_{i}}\right)\times\\
&\frac{1}{V_{\cD}}\int_{\cD}\d\tau_{j}\xi_{j}\left(x_{i+1},\dots,x_{n}\right)M\left(B_{x_{i+1}}\cap B_{x_{i+2}}\cap\dots\cap B_{x_{n}}\right)\\
&=\kappa_{i}\kappa_{j}
\end{align*}
Iterating this procedure in case of multiple terms of $\xi_{i}$ in
the product finally establishes the formal cumulant relation (\ref{eq:formalcumulant}).

So, inserting equations (\ref{eq:defmn}) and (\ref{eq:defxin}) into (\ref{eq:formalcumulant}),
we find that the average Minkowski polynomial of equation (\ref{eq:avMinkstat})
becomes 
\begin{multline}
m:=\frac{\av M}{V_{\cD}}=1-\exp\sum_{n=1}^{\infty}\frac{\left(-\varrho\right)^{n}}{n!V_{\cD}}\int_{\cD}\d\tau_{n}\xi_{n}\left(x_{1},x_{2},\dots,x_{n}\right)\times \\
M\left(B_{x_{1}}\cap B_{x_{2}}\cap\dots\cap B_{x_{n}}\right)\;.
\end{multline}
It can be reconnected to the single MF densities
of the structure by taking the derivative
\[
v_{\mu}=\frac{1}{\alpha_{\mu}}\left.\frac{\partial^{\mu}m\left(t\right)}{\partial t^{\mu}}\right|_{t=0}\;,
\]
which then leads directly to Equation~(\ref{eq:MinkDensDef}), when,
in addition, we use again motion invariance in the exponent to recover
the form of Equation~(\ref{eq:MinkCorrCon}).


\label{lastpage}

\end{document}